\documentstyle[amsfonts,preprint,eqsecnum,aps,prb]{revtex} 
 
\begin{document} 
\title{Functional Characterization of Generalized Langevin Equations} 
\author{Adri\'{a}n A. Budini$\,^{1}$ and Manuel O. C\'aceres$\,^{2}$} 
\address{$\,^{1}$Max Planck Institute for the Physics of Complex Systems,\\ 
N\"{o}thnitzer Str. 38, \\ 
01187 Dresden, Germany} 
\address{$\,^{2}$Centro At\'{o}mico Bariloche, Instituto Balseiro, CNEA, Univ.\\ 
Nac. de Cuyo and CONICET,\\ 
Av. E. Bustillo Km 9.5, 8400 Bariloche, Argentina} 
\date{\today } 
\maketitle 
 
\begin{abstract} 
We present an exact functional formalism to deal with linear 
Langevin equations with arbitrary memory kernels and driven by any noise 
structure characterized through its characteristic functional. No others 
hypothesis are assumed over the noise, neither the fluctuation dissipation 
theorem. We found that the characteristic functional of the linear 
process can be expressed in terms of noise's functional and the Green 
function of the deterministic (memory-like) dissipative dynamics. This 
object allow us to get a procedure to calculate all the Kolmogorov hierarchy 
of the non-Markov process. As examples we have characterized through the 
1-time probability a noise-induced interplay between the dissipative 
dynamics and the structure of different noises. Conditions that lead to 
non-Gaussian statistics and distributions with long tails are analyzed. The 
introduction of arbitrary fluctuations in fractional Langevin equations have 
also been pointed out. 
\end{abstract} 
 
\pacs{PACS numbers: 02.50.Ey, 5.40.Ca, 05.70.Ln} 
 
 
\section{Introduction} 
 
Noise is a basic ingredient of many types of model in physics, mathematics, 
economy, as well as in engineering. Besides that in each area of research
the fluctuations have very different origins, in many cases the evolution 
equation governing the system of interest can be approximated by a suitable 
stochastic differential equation. 
 
In general the driving forces may be any source of fluctuations and the 
system can be characterized by a given potential. The most popular of those 
stochastic differential equations are the one driven by white Gaussian 
fluctuations, therefore the problem can immediately be reduced to the well 
known Fokker-Planck dynamics \cite{Chandrasekar,Kampen,libro}. If the 
fluctuations are not Gaussian we are faced with a problem that is hard to 
solve, but among the different types of fluctuations the so call dichotomic 
noise is a good candidate to study, because in general for any potential 
some conclusions can be drawn \cite{HLe84,West1,Manuel,Annunziato,Masoliver}%
. If the fluctuations are different from the Gaussian ones or the dichotomic 
noise, in general it is not possible to solve the problem for any potential. 
 
In many situations the technical complications of the model can be 
diminished by studying the system in a linear approximation, i.e., around 
the fixed points of the dissipative dynamics. This approximation leads to 
the study of linear stochastic differential equations with arbitrary noises. 
Besides the simplicity of this kind of equations they were the subject of 
numerous theoretical investigation \cite 
{Wang,Heinrichs,Fulinski,Drory,Masoliver1,Porra,Denisov,Srokowski,Srokowski1,Morgado,West2,Jespersen,Caceres2,Caceres1,Picozzi,Stanley,katia99}%
, and also they provide non-trivial models for the study of many different 
mechanism of relaxation in physics, biology and another research areas. 
 
From the previous considerations, it is clear that the usefulness of a 
linear Langevin equation arises from the possibility of working with 
different kind of noise structures. Therefore, one is faced with the 
characterization, in general, of non-Markov processes. These processes can 
only be completely characterized through the whole Kolmogorov hierarchy of 
the stochastic process \cite{Kampen}, i.e.: the knowledge of any $n-$joint 
probability, or equivalently any $n-$time moment and/or cumulant. 
 
In Ref. \cite{Caceres,Budini}, by using a functional technique, we have been 
able to characterize arbitrary linear Langevin equations with local 
dissipation, giving therefore a procedure to calculate the whole Kolmogorov 
hierarchy of the process and any $n-$time moment. Using this previous 
experience, in this paper we are going to generalize our functional 
technique to tackle the more general situation where the dissipative term is 
non-local in time and the noise is also arbitrary; this is what we call a  
{\it Generalized Langevin Equation}, i.e., a linear Langevin equation with 
arbitrary memory and driven by any noise structure  
\begin{equation} 
\frac{d}{dt}u(t)=-\int_{0}^{t}dt^{\prime }\,\Phi \left( t-t^{\prime }\right) 
u(t^{\prime })+\xi (t),\quad u\in (-\infty ,+\infty ),  \label{Intr1} 
\end{equation} 
where $\xi (t)$, the fluctuation term (i.e., the external noise) is 
characterized by its associated functional, and $\Phi \left( t\right) $ is 
an arbitrary memory kernel. 
 
This type of generalized Langevin equation arises quite natural (considering 
Gaussian fluctuations) in the context of the Zwanzig-Mori projector operator 
technique \cite{Zwa61,Mor58}; in this case the fluctuation-dissipation 
theorem is required \cite{Kubo0,Toda}, which imposes that $\Phi \left( 
t\right) =\left\langle \xi (t+\tau )\xi (\tau )\right\rangle /kT$. Therefore 
the dissipative memory must be consistent with the structure of the 
correlation of the Gaussian fluctuations \cite{Kubo0}. Nevertheless, if the 
system is far away from equilibrium, the fluctuation-dissipation theorem is 
not fulfilled and in general the Gaussian assumption is not a good candidate 
to describe the fluctuations of the system. Therefore, in a general 
situation, both the kernel and the noise properties must be considered 
independent elements whose interplay will determine the full stochastic 
dynamics of the process $u(t)$. Along the paper we will be interested in 
characterizing this noise-induced interplay by assuming different kind of 
noises and memory kernels, both in the transient as in the long-time regime. 
 
The paper is organized as follows. In section II, after a short review of 
the functional method, we obtain the characteristic functional for a 
vectorial Langevin equation with memory. In section III we apply this result 
to different situations. First we analyze the case of stable noises; after 
this step, we analyze the stochastic dynamics induced by an exponential 
memory kernel in a process driven by two different fluctuating structures: 
radioactive and Poisson noises. Then the interplay between the kernel and 
the noise properties is emphasized. The necessary noise's properties that 
guarantee a stationary distribution with a long-tail are presented. As an 
example we introduce the Abel noise that induces this asymptotic behavior. 
The characterization of a fractional Langevin equation with arbitrary noise 
is also presented. The interplay between the fractional property of the 
differential operator and the L\'{e}vy noise is analyzed. In section IV we 
give the conclusions. 
 
\section{Functional Characterization of Arbitrary Linear Langevin Equations} 
 
\subsection{The characteristic functional method} 
 
In our previous papers, we have presented a complete characterization of 
Langevin equations with local dissipation by means of the characteristic 
functional of the stochastic process $u(t)$ [$t\in (0,\infty )$]  
\begin{equation} 
G_{u}\left( [k(t)]\right) =\left\langle \exp i%
\textstyle \int%
_{0}^{\infty }dt\,k(t)u(t)\right\rangle .  \label{m2} 
\end{equation} 
Here $k(t)$\ is a test function, and $\left\langle \cdots \right\rangle $\ 
means an average over the stochastic realizations of $u(t)$ belonging to a 
given support. The knowledge of the characteristic functional allows a full 
characterization of the process. In fact, from this functional it is 
possible to calculate the whole Kolmogorov hierarchy, and of course any $n-$ 
time moment and/or cumulant. This follows by defining the $n-$characteristic 
function $G_{u}^{(n)}(\{k_{j},t_{j}\}_{j=1}^{n})$ as  
\begin{equation} 
G_{u}^{(n)}(\{k_{j},t_{j}\}_{j=1}^{n})=G_{u}\left( [k_{\delta }(t)]\right) , 
\label{m3} 
\end{equation} 
where the test function $k_{\delta }(t)$ must be taken as  
\begin{equation} 
k_{\delta }(t)=k_{1}\delta (t-t_{1})+\cdots +k_{n}\delta (t-t_{n}). 
\label{m4} 
\end{equation} 
Thus, any $n-$time joint probability distribution $P(\{u_{j},t_{j}%
\}_{j=1}^{n})\equiv P_{n}\left( u_{1},t_{1};u_{2},t_{2}\cdots 
;u_{n},t_{n}\right) $ can be obtained by Fourier inversion of $%
G_{u}^{(n)}(\{k_{j},t_{j}\}_{j=1}^{n})$  
\begin{equation} 
P(\{u_{j},t_{j}\}_{j=1}^{n})=%
{\displaystyle {1 \over \left( 2\pi \right) ^{n}}}%
\int dk_{1}\cdots \int dk_{n}\,\exp 
(-i\sum\limits_{j=1}^{n}k_{j}u_{j})G_{u}^{(n)}(\{k_{j},t_{j}\}_{j=1}^{n}). 
\label{m5} 
\end{equation} 
On the other hand, any $n-$time moment can be calculated as  
\begin{equation} 
\left\langle u(t_{1})u(t_{2})\cdots u(t_{n})\right\rangle =(-i)^{n}\left.  
\frac{\partial ^{n}G_{u}^{(n)}(\{u_{j},t_{j}\}_{j=1}^{n})}{\partial 
k_{1}\partial k_{2}\cdots \partial k_{n}}\right| _{k_{j}=0}.  \label{m6} 
\end{equation} 
An equivalent formulae holds for the cumulants (or correlation functions), 
in which case the differentiation must be taken from the logarithm of the $%
n- $characteristic function  
\begin{equation} 
\left\langle \left\langle u(t_{1})u(t_{2})\cdots u(t_{n})\right\rangle 
\right\rangle =(-i)^{n}\left. \frac{\partial ^{n}\ln 
G_{u}^{(n)}(\{u_{j},t_{j}\}_{j=1}^{n})}{\partial k_{1}\partial k_{2}\cdots 
\partial k_{n}}\right| _{k_{j}=0}.  \label{cumu} 
\end{equation} 
 
We have showed \cite{Caceres,Budini} that in order to obtain the 
characteristic functional of $u(t)$, it is only necessary to know the 
characteristic functional of the noise $\xi (t)$%
\begin{equation} 
G_{{\bf \xi }}([k(t)])=\left\langle \exp i%
\textstyle \int%
_{0}^{\infty }dt\,k(t)\xi (t)\right\rangle .  \label{m7} 
\end{equation} 
No any other hypothesis were assumed over the noise. This general formalism 
allows us to deal with arbitrary non-Markovian evolution processes, where 
there is not a clear underlying Fokker-Planck dynamics. 
 
In what follows we will generalize our functional characterization for the 
case of a Langevin dynamics that includes memory effects, Eq.~(\ref{Intr1}). 
We remark that not any condition as thermal equilibrium, Gaussian noises, or 
any other, is imposed on the stochastic evolution of the processes $u(t)$. 
 
\subsection{The multivariable linear case with dissipative memory} 
 
A general case that covers many models of evolution, is the vectorial linear 
stochastic equation  
\begin{equation} 
\frac{d}{dt}{\bf u}(t)=-\int_{0}^{t}dt^{\prime }\,{\bf \Phi }\left( 
t-t^{\prime }\right) \cdot {\bf u}(t^{\prime })+{\bf \xi }(t).  \label{g1} 
\end{equation} 
Here, ${\bf u}(t)$ represent a $d-$dimensional stochastic process $\left\{ 
u_{i}(t)\right\} _{i=1}^{d}$, with $u_{i}\in (-\infty ,+\infty )$. The 
memory kernel ${\bf \Phi }\left( t\right) $ is an $d\times d$ matrix with 
arbitrary kernel functions. The $d-$dimensional vector noise ${\bf \xi }(t)$%
\ is characterized through its characteristic functional  
\begin{equation} 
G_{{\bf \xi }}([{\bf k}(t)])=\left\langle \exp i%
\textstyle \int%
_{0}^{\infty }dt\,{\bf k}(t)\cdot {\bf \xi }(t)\right\rangle ,  \label{g2} 
\end{equation} 
where the dot $\{\cdot \}$ denotes a scalar vectorial product, and $\,{\bf k}%
(t)$ is a vector of test functions $\left\{ k_{i}(t)\right\} _{i=1}^{d}$. 
 
We want to found an exact expression for the functional of the process ${\bf %
u}(t)$, defined as  
\begin{equation} 
G_{{\bf u}}([{\bf k}(t)])=\left\langle \exp i%
\textstyle \int%
_{0}^{\infty }dt\,{\bf k}(t)\cdot {\bf u}(t)\right\rangle .  \label{g3} 
\end{equation} 
The basic idea consists in writing the solution of Eq.~(\ref{g1}) in terms 
of the corresponding Green function. By denoting the Laplace transform as $%
\tilde{f}(s)=\int_{0}^{\infty }e^{-st}f(t)\,dt$, from Eq.~(\ref{g1}) we get  
\begin{equation} 
s{\bf \tilde{u}}(s)-{\bf u}(0)=-{\bf \tilde{\Phi}}\left( s\right) \cdot {\bf  
\tilde{u}}(s)+{\bf \tilde{\xi}}(s).  \label{g4} 
\end{equation} 
This allows to express the formal solution for each realization of the 
noises as  
\begin{equation} 
{\bf u}(t)=\left\langle {\bf u}(t)\right\rangle _{0}+\int_{0}^{t}dt^{\prime 
}\,{\bf \Lambda }\left( t-t^{\prime }\right) \cdot {\bf \xi }(t^{\prime }), 
\label{g5} 
\end{equation} 
where we have defined the vector  
\begin{equation} 
\left\langle {\bf u}(t)\right\rangle _{0}\equiv {\bf \Lambda }\left( 
t\right) \cdot {\bf u}(0).  \label{g6} 
\end{equation} 
The $d\times d$ matrix ${\bf \Lambda }\left( t\right) $ is defined through 
its Laplace transform  
\begin{equation} 
{\bf \tilde{\Lambda}}\left( s\right) =\frac{1}{s\text{{\bf \^{I}}}+{\bf  
\tilde{\Phi}}\left( s\right) },  \label{g7} 
\end{equation} 
where {\bf \^{I}}\ is a $d-$dimensional identity matrix and ${\bf \tilde{\Phi%
}}\left( s\right) $ is the Laplace transform of the matrix kernel ${\bf \Phi  
}\left( t\right) $. Eq.~(\ref{g7}) is equivalent to the evolution  
\begin{equation} 
\frac{d}{dt}{\bf \Lambda }(t)=-\int_{0}^{t}dt^{\prime }\,{\bf \Lambda }%
(t^{\prime })\cdot {\bf \Phi }\left( t-t^{\prime }\right) ,\quad {\bf %
\Lambda }(0)=1,  \label{g8} 
\end{equation} 
From this equation it is possible to identify the matrix ${\bf \Lambda }(t)$ 
with the Green function of Eq.~(\ref{g1}). Notice that the vector $%
\left\langle {\bf u}(t)\right\rangle _{0}$ corresponds to the average value 
of the process ${\bf u}(t)$ for the case in which the average of the noise 
is null, i.e., $\left\langle {\bf \xi }(t)\right\rangle =0$. 
 
After introducing the solution Eq.~(\ref{g5}) into Eq.~(\ref{g3}), and 
reordering the time integrals $\int_{0}^{t}dt^{\prime }\int_{0}^{t^{\prime 
}}dt^{\prime \prime }=\int_{0}^{t}dt^{\prime \prime }\int_{0}^{t}dt^{\prime 
}\Theta (t^{\prime }-t^{\prime \prime })$, where $\Theta (t)$ is the step 
function, we arrive to a closed expression for the characteristic functional 
of ${\bf u}(t)$ in terms of the functional of the vector noise. Thus, we get  
\begin{equation} 
G_{{\bf u}}([{\bf k}(t)])=G_{\left\langle {\bf u}\right\rangle _{0}}([{\bf k}%
(t)])\,G_{{\bf \xi }}([{\bf z}(t)]),  \label{g9} 
\end{equation} 
where we have defined  
\begin{equation} 
G_{\left\langle {\bf u}\right\rangle _{0}}([{\bf k}(t)])=\exp \{i%
\textstyle \int%
_{0}^{\infty }dt\,{\bf k}(t)\cdot \left\langle {\bf u}(t)\right\rangle 
_{0}\},  \label{g10} 
\end{equation} 
and the vector test function ${\bf z}(t)$ as  
\begin{equation} 
{\bf z}(t)=\int_{t}^{\infty }dt^{\prime }\,{\bf k}(t^{\prime })\cdot {\bf %
\Lambda }(t^{\prime }-t).  \label{g11} 
\end{equation} 
 
The expression Eq.~(\ref{g9}) gives us the desired exact functional $G_{{\bf %
u}}([{\bf k}(t)])$ as the product of two functional. The first one 
corresponds to the deterministic evolution, or equivalently to the averaged 
process when the noises have null averages. On the other hand, the second 
term comes from the noise characteristic functional, i.e., the stochastic 
evolution. 
 
Finally, the $n-$time characteristic function $G_{{\bf u}}^{(n)}(\{{\bf k}%
_{j},t_{j}\}_{j=1}^{n})$ can immediately be evaluated from the 
characteristic functional of the noise. The unidimensional case, Eq.~(\ref 
{m3}), is easily generalized to the vectorial case as  
\begin{equation} 
G_{{\bf u}}^{(n)}(\{{\bf k}_{j},t_{j}\}_{j=1}^{n})=G_{{\bf u}}([{\bf k}%
_{\delta }(t)]),  \label{g12} 
\end{equation} 
where the vectorial test function ${\bf k}_{\delta }(t)$ is  
\begin{equation} 
{\bf k}_{\delta }(t)={\bf k}_{1}\delta (t-t_{1})+\cdots +{\bf k}_{n}\delta 
(t-t_{n}).  \label{g13} 
\end{equation} 
Thus, using these last two equations and Eq.~(\ref{g9}) to (\ref{g11}), we 
get  
\begin{equation} 
G_{{\bf u}}^{(n)}(\{{\bf k}_{j},t_{j}\}_{j=1}^{n})=\exp \{i%
\mathop{\textstyle \sum }%
\nolimits_{j=1}^{n}{\bf k}_{j}\cdot \left\langle {\bf u}(t_{j})\right\rangle 
_{0}\}\,G_{{\bf \xi }}([{\bf y}(t)]),  \label{g14} 
\end{equation} 
where the function ${\bf y}\left( t\right) $ reads  
\begin{equation} 
{\bf y}\left( t\right) =\sum\nolimits_{j=1}^{n}\Theta (t_{j}-t)\,{\bf k}%
_{j}\cdot {\bf \Lambda }(t_{j}-t).  \label{g15} 
\end{equation} 
 
Eq.~(\ref{g14}) gives, in a simple way, the $n-$characteristic function of 
the process ${\bf u}(t)$ by evaluating the characteristic functional of the 
noise with the function ${\bf y}\left( t\right) $. This last function is 
defined in terms of the Green function ${\bf \Lambda }(t)$ of the problem. 
At this point, it is important to remark that our formalism is valid 
independently of the assumed form for the dissipative Green function 
evolution. Thus, this function may present a monotonous or oscillating decay 
(taking positive and negative values), and in general any one consistent 
with the corresponding kernel. In Appendix A we show a particular $2-$%
dimensional stochastic process $\left\{ u_{i}(t)\right\} _{i=1}^{2}$ where 
each component $u_{i}(t)$ represents the position and the velocity of a 
memory damped harmonic oscillator in presence of an arbitrary noise 
structure. 
 
\section{Examples} 
 
Here we are going to analyze, and characterize with our functional method, 
many amazing situations that arise by choosing different noise structures 
and a non-local dissipative kernel. 
 
\subsection{Local Dissipation} 
 
From now on, we will restrict to a unidimensional case. By assuming a $%
\delta -$Dirac correlated kernel  
\begin{equation} 
\Phi \left( t\right) =\gamma \,\delta \left( t\right) ,  \label{local1} 
\end{equation} 
we arrive to the evolution 
 
\begin{equation} 
\frac{d}{dt}u(t)=-\gamma \,u(t)+\xi (t).  \label{local2} 
\end{equation} 
Therefore, it is possible to obtain $\tilde{\Lambda}\left( s\right) 
=1/(s+\gamma )$. Thus,  
\begin{equation} 
\Lambda \left( t\right) =\exp \left[ -\gamma \,t\right] .  \label{local3} 
\end{equation} 
With this solution, it is possible to recapture all the results obtained in 
Ref. \cite{Caceres}. 
 
\subsection{Stable noises} 
 
Stable probability distributions play an important role in the theory of 
sums of random variables. This fact follows from the generalized central 
limit theorem valid for L\'{e}vy distributions and the asymptotic power-law 
decay \cite{Levy,GKo54}. Here we are going to apply our functional formalism 
with the use of stable noises. 
 
\subsubsection{Gaussian noise} 
 
A zero-mean Gaussian noise $\xi (\tau ),$ with an arbitrary correlation 
function $\sigma _{\xi }(\tau _{2},\tau _{1})=\left\langle \xi (\tau 
_{2})\xi (\tau_{1})\right\rangle ,$ is characterized by the functional \cite 
{Kampen}  
\begin{equation} 
G_{\xi }([k(t)])=\exp \left( -\frac{1}{2}\int_{0}^{\infty }d\tau 
_{2}\int_{0}^{\infty }d\tau _{1}\,k(\tau _{2})\sigma _{\xi }(\tau _{2},\tau 
_{1})k(\tau _{1})\right) .  \label{gaus1} 
\end{equation} 
Therefore, from Eqs.~(\ref{g9}) to (\ref{g11}) the characteristic functional 
of the process $u(t)$ results  
\begin{equation} 
G_{u}([k(t)])=G_{\left\langle u\right\rangle _{0}}([k(t)])\,\exp \left( -%
\frac{1}{2}\int_{0}^{\infty }d\tau _{2}\int_{0}^{\infty }d\tau _{1}\,k(\tau 
_{2})k(\tau _{1})\sigma _{u}(\tau _{2},\tau _{1})\right) , 
\end{equation} 
where  
\begin{equation} 
\sigma _{u}(\tau _{2},\tau _{1})=\int_{0}^{\tau _{2}}d\tau_{b} 
\int_{0}^{\tau _{1}}d\tau_{a} \,\Lambda (\tau _{2}-\tau_{b} )\sigma _{\xi 
}(\tau_{b} ,\tau_{a} )\Lambda (\tau _{1}-\tau_{a} ),  \label{gaus3} 
\end{equation} 
is the correlation function of the Gaussian process $u(t)$. In fact, the $n-$%
time characteristic function of $u(t)$ is a multivariate Gaussian process. 
 
\subsubsection{L\'{e}vy noise} 
 
Another example that share similar conclusions than using the Gaussian 
noise, is the L\'{e}vy noise. The characteristic functional of this noise 
reads\cite{Caceres2}  
\begin{equation} 
G_{\xi }([k(t)])=\exp \left( -\frac{\sigma _{\nu }}{2}\int_{0}^{\infty 
}d\tau \left| k(\tau )\right| ^{\nu }\,\right) ,  \label{levy1} 
\end{equation} 
where $0<\nu <2$. Thus, for example, the $1-$time characteristic function of 
the process $u(t)$ reads  
\begin{equation} 
G_{u}^{(1)}(k,t)=\exp \{ik\left\langle u(t)\right\rangle _{0}\}\,\exp \left( 
-\frac{1}{2}\left| k\right| ^{\nu }\Xi _{u}(t)\right) ,  \label{ulevy} 
\end{equation} 
where  
\begin{equation} 
\Xi _{u}(t)=\sigma _{\nu }\int_{0}^{t}dt^{^{\prime }}\left| \Lambda 
(t^{^{\prime }})\right| ^{\nu }.  \label{Levymemo} 
\end{equation} 
Therefore, memory-like linear Langevin equations driven by stable noises 
give rise to new stable stochastic processes whose correlations properties 
are defined in terms of the Green function $\Lambda (t)$ of the dissipative 
dynamics. Note that this result is valid independently of the particular 
form of the Green function. 
 
\subsection{Interplay between non-local friction and noise structure} 
 
In general, around a fixed point, the linearized dynamics of a dissipative 
system---depending on the parameters of the problem---may presents a 
relaxation behavior with quite different characteristics: let us say, a 
monotonous or a time-oscillating decay. Here we are going to study the 
relaxation and the fluctuations of the process $u(t)$ by analyzing the 
interplay between these mentioned dissipative behaviors and different noise 
structures. 
 
The different behaviors of the dissipative dynamics can be modeled through a 
non-local exponential memory friction  
\begin{equation} 
\Phi \left( t\right) =\delta \exp \left[ -\lambda t\right] ,  \label{exp1} 
\end{equation} 
therefore $\tilde{\Phi}\left( s\right) =\delta /\left( s+\lambda \right) $.  
From Eq.~(\ref{g7}) we get $\tilde{\Lambda}(s)=(s+\lambda )/[s(s+\lambda 
)+\delta ],$ which after inversion gives  
\begin{equation} 
\Lambda \left( t\right) =e^{-\lambda t/2}\left\{ \cos \left[ w_{0}t\right] +%
\frac{\lambda }{2w_{0}}\sin \left[ w_{0}t\right] \right\} ,  \label{exp2} 
\end{equation} 
where the frequency $w_{0}$ is  
\begin{equation} 
w_{0}=\sqrt{\delta -\left( \frac{\lambda }{2}\right) ^{2}}.  \label{complex} 
\end{equation} 
As desired, the dissipative Green function, Eq.(\ref{exp2}), has a regime 
where it decays oscillating in time. But on the other hand there is a 
possibility, for $\delta <\left( \frac{\lambda }{2}\right) ^{2}$ where $%
\Lambda \left( t\right) $ decays in a monotonous way. Note that the local 
case (non-memory dissipation), Eq.~(\ref{local1}), is re-obtained 
considering the limit: $\lambda \rightarrow \infty $ and $\delta \rightarrow 
\infty ,$ with $\delta /\lambda \rightarrow \gamma $ (finite). 
 
Now we are going to present some results using the Green function Eq.~(\ref 
{exp2}) in connection with two different structures for the fluctuating 
term, i.e., a radioactive noise and a Poisson white-noise. We will show that 
the interplay between the deterministic Green function and the structure of 
the noise play a crucial role to determine the relaxation and in general any 
statistical dynamical property of the stochastic process $u(t).$ On the 
other hand, a memory kernel with a power-law decay will be studied in the 
context of fractional derivatives in the sub-section E. 
 
\subsubsection{Radioactive decay noise} 
 
The radioactive noise is non-white and Markovian. Its $1-$time and 
conditional probability evolve controlled by the master equation  
\begin{equation} 
\frac{dP_{\xi }(t)}{dt}=\beta \,\,\left[ (\xi +1)P_{\xi +1}(t)-\xi \,P_{\xi 
}(t)\right] . 
\end{equation} 
Thus, the realizations of this noise start with some initial value $\xi 
_{0} $ and at random times decrease by finite unit steps until the process 
reaches the zero value. The constant $\beta $ defines the probability per 
unit of time for such discrete steps ($\beta -$decay). The characteristic 
functional reads \cite{Kampen}  
\begin{equation} 
G_{\xi }([k(t)])=\left( \beta \int_{0}^{\infty }d\tau \,\exp \left[ -\beta 
\tau +i\int_{0}^{\tau }d\tau ^{\prime }k(\tau ^{\prime })\right] \right) 
^{\xi _{0}}.  \label{rad1} 
\end{equation} 
This noise clearly does not reach a stationary regime. Using this noise, the  
$1-$ time characteristic function of the process $u(t)$ results  
\begin{equation} 
G_{u}^{(1)}(k,t)=e^{ik\left\langle u(t)\right\rangle _{0}}\,\left( e^{-\beta 
t} \exp \left[ ik\int_{0}^{t}d\tau ^{\prime }\Lambda (\tau ^{\prime 
})\right] +\beta \int_{0}^{t}d\tau \,e^{-\beta \tau }\exp \left[ 
ik\int_{t-\tau }^{t}d\tau ^{\prime }\,\Lambda (\tau ^{\prime })\right] 
\right) ^{\xi _{0}}.  \label{nuclear} 
\end{equation} 
 
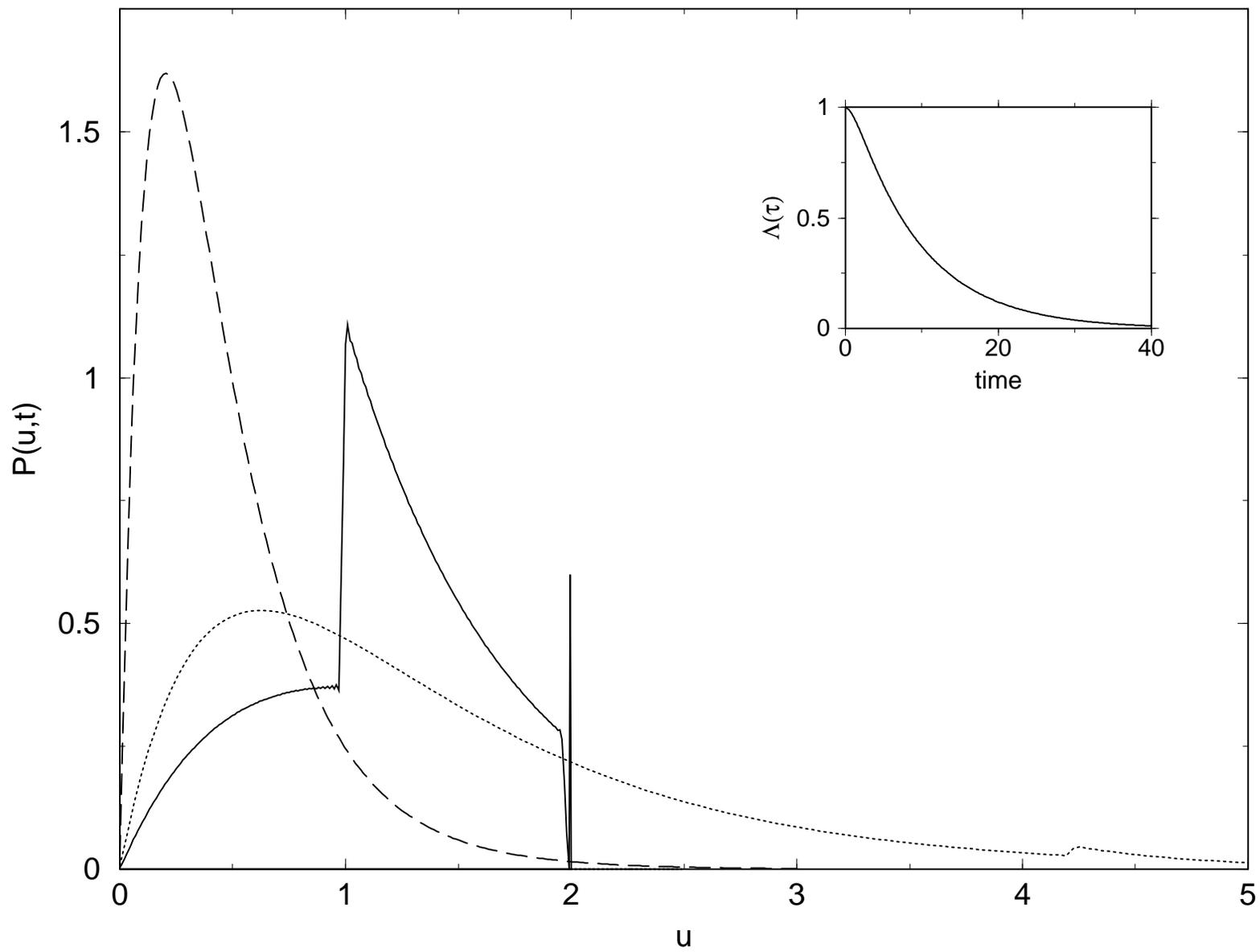
\begin{figure}[tbp] 
\caption{ 
Probability distribution, $P(u,t),$ for a memory-like driven process $u(t)$ 
with an exponential dissipative kernel and in presence of a Radioactive 
noise, $\xi (t),$ as a function of $u$ for three different times, $t=1$ 
(full line), $t=5$ (dotted line), $t=15$ (dashed line), taken in arbitrary 
units. The initial condition was chosen as $u(t=0)=0$. The noise parameters 
are $\beta =1$ and $\xi _{0}=2$.  The parameters of the Green function are $%
\delta =0.1$ and $\lambda =1$; the inset shows its monotonous decaying 
behavior as a function of time. The straight line in $P(u,t=1)$ indicates 
the $\delta -$Dirac contribution (at short times) in the distribution of the 
process $u(t)$. Asymptotically $P(u,t\rightarrow \infty )$ goes to a $\delta 
-$Dirac located at $u=0.$ 
} 
\label{radio1}  
\end{figure}

 
This noise leads to very striking dynamical behaviors for the driven process  
$u(t)$. In the following figures we have plotted, at different times, the 
corresponding exact probability distribution $P(u,t)$. In order to get this 
object we have anti-transformed Eq.~(\ref{nuclear}) by using a fast Fourier 
algorithm. In Fig.~\ref{radio1}, the parameters for the memory kernel were 
chosen as $\delta =0.1$, $\lambda =1$. Thus, the Green function decays 
without oscillations (see inset). The noise decay rate is $\beta =1$ and its 
initial value was chosen to be $\xi _{0}=2$. At time $t=1$ (full line), we 
see that the probability distribution is highly irregular and also has a $%
\delta -$Dirac contribution (indicated symbolically with the vertical line). 
This singular behavior arises from the particular realization $\xi (t)$ in 
which the noise has not yet suffer any decay. 
 
Due to the non-vanishing value of the noise average, $\left\langle \xi 
(t)\right\rangle \neq 0,$ the $\delta -$Dirac contribution has different 
locations at different times. This term can be directly read from the 
characteristic functional of the process $u(t)$, Eq.~(\ref{nuclear}), which 
gives: $\exp \left( -\xi _{0}\beta t\right) \,\delta [u-\xi 
_{0}\int_{0}^{t}d\tau ^{\prime }\,\Lambda (\tau ^{\prime })]$. The 
discontinuity of the probability distribution $P(u,t)$ is due to the strong 
change that arises in the driving noise after a $\beta -$decay. In fact, in 
this example, the noise intensity decreases to half of its initial value. At 
a later time, $t=5$ (dotted line), due to the dissipative dynamics the 
probability distribution $P(u,t)$ loss it discontinuous character and seems 
to be accumulated near of the origin. At even later time, $t=15$ (dashed 
line) this accumulation seems to increase. As a matter of fact, the 
stationary distribution is a $\delta -$Dirac centered in $u=0$. This 
behavior is a consequence of the coupling between the noise properties and 
the deterministic dynamics of $u(t)$: as the noise intensity dead at long 
times, all realizations $u(t)$ are attracted by the stable point of the 
dissipative dynamics, which corresponds to $u=0$. 
 
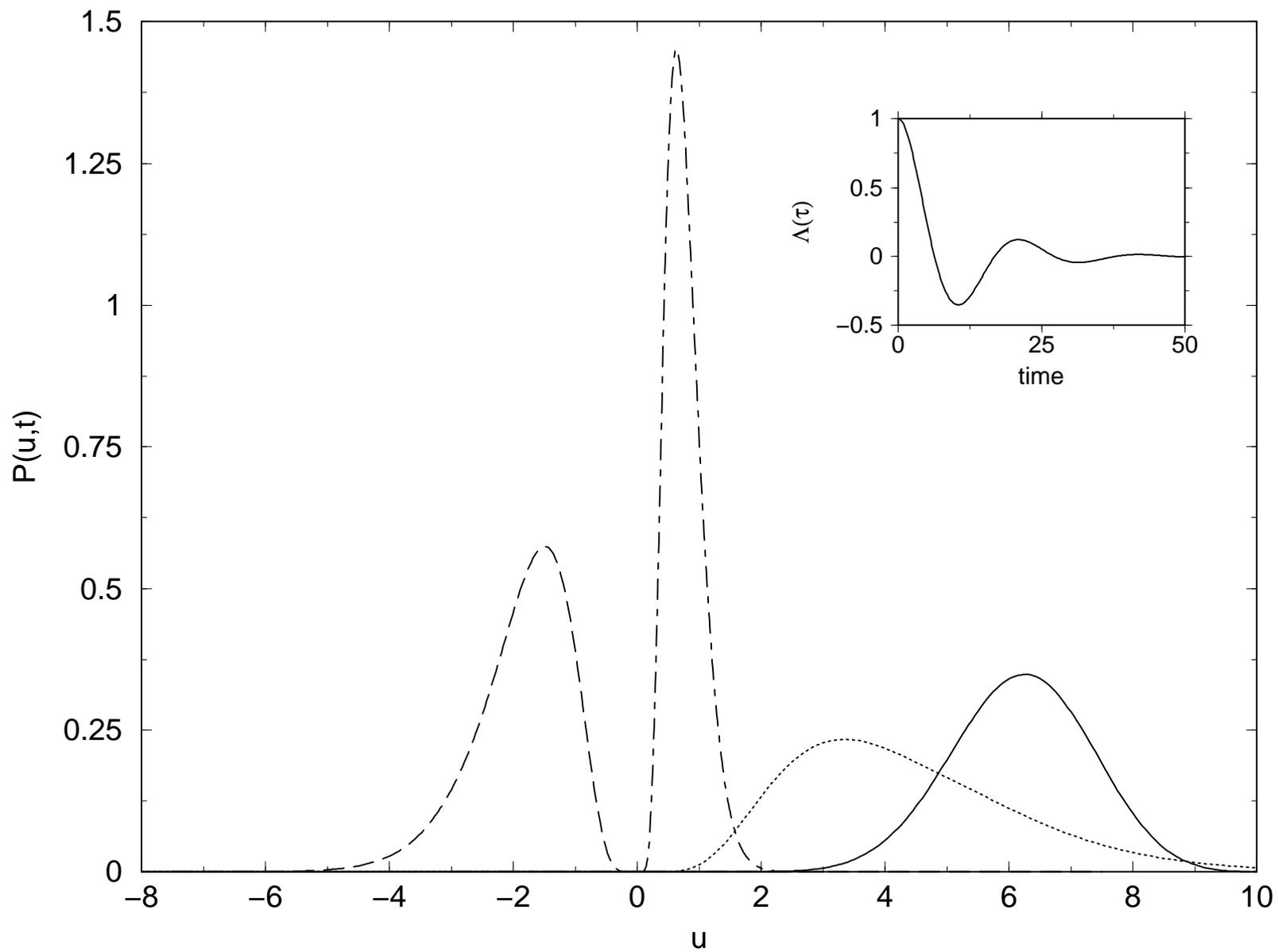
\begin{figure}[tbp]  
\caption{  
Probability distribution $P(u,t)$ as in Fig.~\ref{radio1} for four different 
times, $t=1$ (continuous line), $t=5$ (dotted line), $t=15$ (dashed line), $%
t=25$ (dot-dashed line), in arbitrary units. The noise parameters are $\beta 
=1$ and $\xi _{0}=10$. The parameters of the Green function are $\delta =0.1$ 
and $\lambda =0.2$; the inset shows its oscillatory decaying behavior as a 
function of time. Here, at large times the distribution $P(u,t\rightarrow 
\infty )$ also goes asymptotically to a $\delta -$Dirac located at $u=0.$  
} 
\label{radio2}  
\end{figure}

 
In Fig.~\ref{radio2} we show the case in which the Green function is 
oscillatory in time (see inset). The parameters are $\delta =0.1$, $\lambda 
=0.2$, and for the noise we choose $\beta =1$ and $\xi _{0}=10$. Here, in 
contrast to the previous figure, we have increased the initial intensity of 
the noise, which imply that the $\delta -$Dirac contribution decay in a very 
fast way. On the other hand, this higher initial value imply that the 
``discrete'' decay of the corresponding noise realizations do not change 
appreciably the noise intensity. In consequence, at all time the 
distribution $P(u,t)$ results smooth. As in the previous example, all 
realization of the process $u(t)$ are attracted by the stable point $u=0$. 
Nevertheless, here the transient behavior reflects the oscillatory behavior 
of the Green function. In fact, we note that at successive times, the {\it %
center of mass} of the distribution $P(u,t)$ approximately follows the 
dissipative dynamics of the Green function oscillating around the origin $%
u=0 $ (see times: $t=1$ continuous line, $t=5$ dotted line, $t=15$ dashed 
line, $t=25$ dot-dashed line). In addition, we note that the width of the 
distribution grows approximately up to $t=5$. In fact, at this time most of 
the noise realizations have decayed to a null value. For later times, the 
distribution is mainly governed by the dissipative dynamics of $u(t)$. 
Therefore, its width decreases up to reaching a $\delta -$Dirac form in the 
stationary state. 
 
We remark that for this noise the stationary distribution $P(u)$ is always 
determinate by the fixed point of the dissipative dynamics. This 
characteristic is due to the vanishing noise's amplitude at the long time 
regime. In the next example we show a case where the noise properties 
participate, in a crucial way, both in the transient as in the stationary 
properties of the process $u(t)$. 
 
\subsubsection{Shot noise} 
 
The stochastic realizations of the shot noise are defined by a sequence of 
pulses $\psi (t)$, each one arriving at random independent times $t_{i}$. 
The characteristic functional of the noise \cite{Kampen,Caceres,Budini} $%
{\bf \xi }(t)$, in the interval $t\in [0,\infty ]$, is  
\begin{equation} 
G_{\xi }([k(t)])=\exp \left( \int_{0}^{\infty }d\tau \,\,q(\tau )\left[ \exp 
\left( i\int_{0}^{\infty }k(t)\psi (t-\tau )\,dt\right) -1\right] \right) , 
\end{equation} 
where $q(\tau )$ is the density of arriving pulses. Therefore, the complete 
characterization of the process $u(t)$ is given by  
\begin{equation} 
G_{u}([k(t)])=G_{\left\langle u\right\rangle _{0}}([k(t)])\exp \left( 
\int_{0}^{\infty }d\tau \,q(\tau )\left[ \exp \left( i\int_{0}^{\infty 
}k(t)\Psi (t,\tau )\,dt\right) -1\right] \right) ,  \label{Cambel1} 
\end{equation} 
where the function $\Psi (t,\tau )$ is defined as  
\begin{equation} 
\Psi (t,\tau )=\int_{0}^{t}dt^{\prime }\,\Lambda (t-t^{\prime })\psi 
(t^{\prime }-\tau ). 
\end{equation} 
If $\psi (t)=A\delta (t)$ and the density of ``dots'' $q(\tau )$ is uniform $%
q(\tau )=\rho $ the noise ${\bf \xi }(t)$ reduces to the more familiar white 
shot-noise or Poisson noise whose characteristic functional is  
\begin{equation} 
G_{\xi }([k(t)])=\exp \left( \rho \int_{0}^{\infty }dt\left[ \exp 
[iAk(t)]-1\right] \right) .  \label{shot2} 
\end{equation} 
By adding two statistically independent white shot-noises, with opposite 
amplitudes, we get  
\begin{equation} 
G_{\xi }([k(t)])=\exp \left( 2\rho \int_{0}^{\infty }dt\left[ \cos \left( 
Ak(t)\right) -1\right] \right) .  \label{poisonsimetrico} 
\end{equation} 
This functional characterizes a noise with a null average, whose 
realizations consists in the random arriving of $\delta -$Dirac pulses with 
amplitude $\pm A$. Note that in the limit $A\to 0$, $\rho \to \infty $ with $%
A^{2}\rho =D/2$, this symmetrical white shot-noise converges to a Gaussian 
white noise, i.e., Eq.~(\ref{gaus1}) with $\sigma _{\xi }(\tau _{2},\tau 
_{1})=D\delta (\tau _{2}-\tau _{1})$. 
 
Using Eq.~(\ref{poisonsimetrico}), the $1-$time characteristic function of 
the process $u(t)$ is given by  
\begin{equation} 
G_{u}^{(1)}(k,t)=\exp \{ik\left\langle u(t)\right\rangle _{0}\}\,\exp \left( 
2\rho \int_{0}^{t}d\tau \,\left[ \cos \left( Ak\Lambda (\tau )\right) 
-1\right] \right) .  \label{simetricol} 
\end{equation} 
As we will show below, this expression leads to a rich variety of possible 
stochastic dynamical behavior for the process $u(t)$; both, in the transient 
as in the stationary state, the different behaviors arise from the 
competence between the different characteristic time scales of the Green 
function [Eq.~(\ref{exp2})] and those of the noise. In the monotonous 
regime, $\delta <\lambda ^{2}/4$, the decay of $\Lambda (\tau )$ can be 
characterized by the rate $\delta /\lambda $. For $\delta >\lambda ^{2}/4$, 
when $w_{0}>\lambda /2$, the oscillatory behavior of $\Lambda (\tau )$ is 
characterized by the damping rate $\lambda /2$ and the frequency $w_{0}$. On 
the other hand, the noise is characterized the rate $\rho $ and the 
amplitude $A$. The analysis of the behavior of $P(u,t)$ on these parameters 
can be simplified by defining the rescaled process $u^{^{\prime }}(t)=u(t)/A$%
, whose evolution can be written as  
\begin{equation} 
\frac{du^{^{\prime }}(\tau ^{^{\prime }})}{d\tau ^{^{\prime }}}=-\delta {%
^{\prime }}\int_{0}^{\tau ^{^{\prime }}}d\tau ^{^{\prime \prime }}\,\exp 
\left[ -(\tau ^{^{\prime }}-\tau ^{^{\prime \prime }})\right] \,u^{^{\prime 
}}(\tau ^{^{\prime \prime }})+\xi ^{^{\prime }}(\tau ^{^{\prime }}). 
\end{equation} 
Here, we have defined the dimensionless time $\tau ^{^{\prime }}=\lambda t$ 
and parameter $\delta {^{\prime }}=\delta /\lambda ^{2}$. Furthermore, the 
dimensionless symmetric Poisson noise $\xi ^{^{\prime }}(\tau ^{^{\prime }})$ 
is only characterized by the dimensionless rate $\rho ^{^{\prime }}=\rho 
/\lambda $. Thus, the full dynamical properties of the process $u(t)$ are 
controlled by the parameters $\delta ^{^{\prime }}$ and $\rho ^{^{\prime }}$. 
 
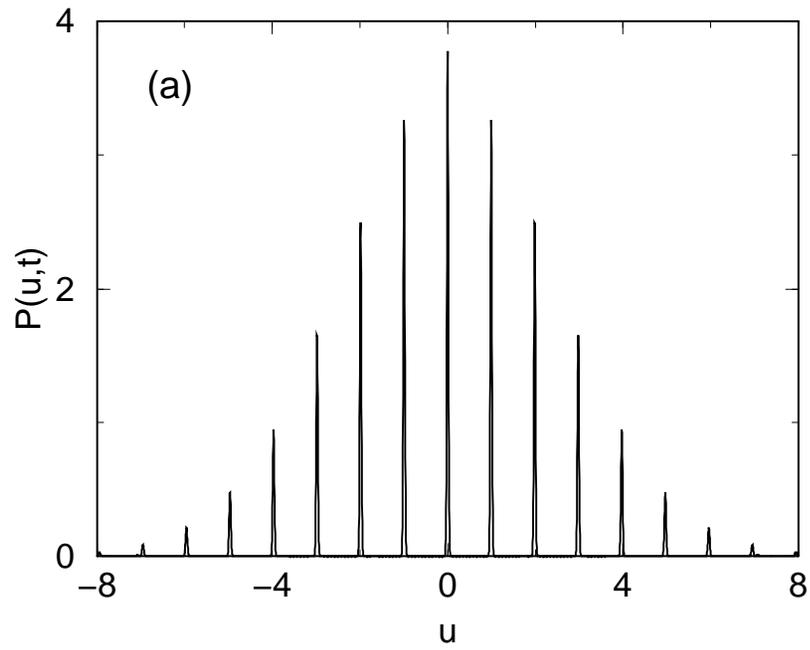
\begin{figure}[tbp]  
\caption{  
Probability distribution, $P(u,t),$ for a memory-like driven process $u(t)$ 
with an exponential dissipative kernel and in presence of a symmetric white Poisson 
noise, $\xi (t),$ as a function of $u$ for five different times, Fig.~\ref 
{poison1}(a) $t=0.5$, Fig.~\ref{poison1}(b) $t=1.65$, Fig.~\ref{poison1}(c) $
t=2$, Fig.~\ref{poison1}(d) $t=2.5$ (dotted line) and $t=20$ (full line), 
taken in arbitrary units. The initial condition was chosen as $u(t=0)=0.$  
The noise parameters are $A=1$, $\rho =7.5$. The parameters of the Green 
function are $\delta =0.2$, $\lambda =1$; the inset in Fig.~\ref{poison1}(d) 
shows its monotonous decaying behavior as a function of time. } 
\label{poison1}  
\end{figure}

 
In the next figures we have obtained numerically $P(u,t)$ from fast Fourier 
transform of Eq.~(\ref{simetricol}). In Fig.~\ref{poison1} the parameters of 
the kernel are $\delta =0.2$, $\lambda =1$, and for the noise we used $A=1$,  
$\rho =7.5$. For these values, the frequency $w_{0}$ in Eq.~(\ref{complex}) 
is complex. Therefore, the Green function $\Lambda (t)$ decays monotonously 
in time (see inset). At short times, $t=0.5$ [Fig.~\ref{poison1}(a)], the 
probability distribution presents a series of peaks, each ones separated by 
a distance $A=1$. This property is a direct consequence of the nature of the 
shot-noise, where each arriving $\delta -$Dirac pulse produces a shift of 
magnitude $A$ in the driven process $u(t)$. In this short time regime the 
effect of the dissipative dynamics is negligible. At later times, the 
dissipative contribution becomes appreciable; its effects is to attract all 
the realization of $u(t)$ to the fixed point $u=0$. This action, added to 
the shift effect of the arriving pulses, produces an increasing of the width 
of all the peaks of the distribution; then any value of $u$ becomes 
(approximately) probable. This effect is clearly seen in Fig.~\ref{poison1}%
(b) [$t=1.65$]. Note that the distribution retained the ``sharpy'' structure 
of peaks. This erasing effect is even increased with time; see Fig.~\ref 
{poison1}(c) (time $t=2$). At later times, all the signature of the peaks 
have disappeared and the distribution goes asymptotically to a Gaussian 
distribution; see Fig.~\ref{poison1}(d) [times $t=2.5$ (dotted line) and $%
t=20$ (full line)]. 
 
In general, an asymptotic Gaussian stationary distribution is always 
expected when the rate of the arriving $\delta -$Dirac pulses is greater 
than the characteristic decay rate of the Green function. When the decay of $%
\Lambda (t)$ is monotonous, this condition can be expressed through the 
inequality $\rho ^{^{\prime }}>\delta ^{^{\prime}}$, which is clearly 
satisfied in the previous figure. When this relation is not satisfies, 
deviations from a Gaussian distribution must arise both in the transient as 
in the stationary distribution. We show this effect in the next figures. 
 
 
In Fig.~\ref{poison2}, the parameters of the Green function (see inset) are 
the same than in the previous figure, [$\delta =0.2$, $\lambda =1$], and for 
the noise we have used $A=1$, $\rho =0.1$. As in the previous case, the 
behavior of $P(u,t)$ arises from the competence between the shift effect of 
the arriving pulses and the dissipative dynamics of $u(t)$. For example, at 
short times [Fig.~\ref{poison2}(a), $t=2.5$] the probability shows a set of 
maxima located at $\pm A$ with tails pointing in the direction to the 
origin. These tails are mainly originated by the dissipative dynamics that 
attracts all realizations to the stable point $u=0$. On the other hand, here 
the occupation around $u=0$ comes from the shift effect produced by two 
arriving pulses with different signs. In contrast to the previous example, 
at later times [Fig.~\ref{poison2}(b), $t=5$] the evolution of the 
distribution $P(u,t)$ is mainly governed by the dissipative dynamics and the 
structure of peaks is retained during all the evolution, Fig.~\ref{poison2}%
(c) $t=10$, and Fig.~\ref{poison2}(d) $t=100$, where the stationary state is 
practically reached. Note that in this stationary state, the non-Gaussian 
characteristics are present in $u\approx 0$ and in $u\approx \pm A$. 
 
We note that during all the transient dynamics, there exist a $\delta $%
-Dirac contribution (indicated with the vertical lines) centered at the 
origin, $u=0$. This term is originated by the noise realization in which not 
any pulse has arrived up to time $t$. Therefore, it is exponentially damped 
with a rate $2\rho $, i.e., it can be written as $\exp \left( -2\rho 
t\right) \delta (u)$. At each of the chosen times, the weight of the $\delta 
-$Dirac term are respectively: $0.606$, $0.368$, $0.135$ and $2\times 
10^{-9} $. Notice that in the stationary state, the $\delta -$Dirac 
contribution is always completely washed out. We remark that, independently 
of the values of the parameters, this $\delta -$Dirac contribution is always 
present. Nevertheless, note that in the case of Fig.~\ref{poison1}, it is 
rapidly attenuated and its contribution is insignificant in the time scale 
of that plot. 
 
In the two previous figures we have assumed a monotonous decreasing Green 
function. In general the interplay between an asymmetric Poisson noise, Eq.~(%
\ref{shot2}), and an oscillatory Green function is similar to that founded 
in the case of the Radioactive noise. In the case of a symmetric noise, Eq.~(%
\ref{simetricol}), due to the symmetry of the problem, the transient 
dynamics is approximately similar to that founded for a monotonous decay. 
Thus, the more relevant aspect to analyze is the corresponding stationary 
distribution. 
 
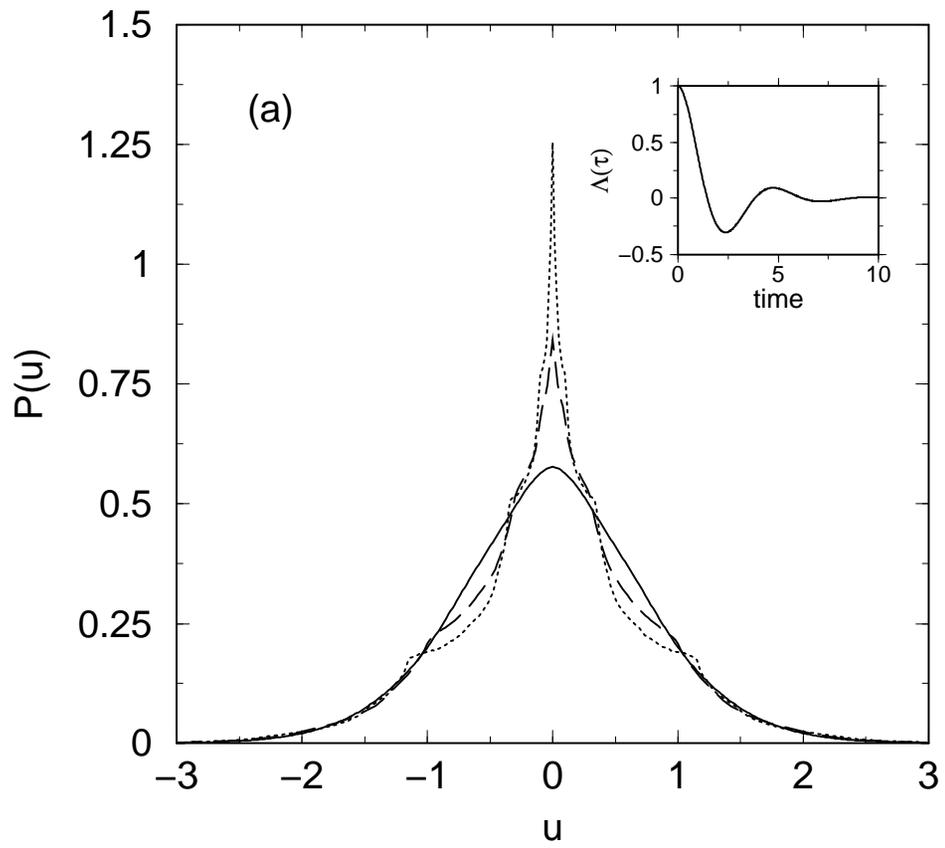
\begin{figure}[tbp]  
\caption{ 
Stationary probability distribution, $P(u),$ for a memory-like driven 
process $u(t)$ with an exponential dissipative kernel and in presence of a 
symmetric white Poisson noise, as a function of $u$ for different values of the 
parameters for the Green function and the noise. In Fig.~\ref{poison3}(a) $%
\delta =2$, $\lambda =1,$ while for the Poisson noise we take $\rho =0.3$ 
(dotted line), $\rho =0.4$ (dashed line) and $\rho =0.8$ (continuous line) 
and $A=\sqrt{0.4/\rho }$. In Fig.~\ref{poison3}(b) the noise parameters are 
the same in each plot, $\rho =0.25$, $A=1$, but we change the parameters for 
the Green function, $\lambda =1$, $\delta =0.15$ (full line), $\delta =0.6$ 
(dashed line), and $\delta =50$ (dotted line). The insets show the Green 
functions for the corresponding different parameters. 
} 
\label{poison3}  
\end{figure}

 
In Fig.~\ref{poison3} we study the properties of the stationary distribution  
$P(u)$ for different values of the parameter of the Green function and the 
noise. In Fig.~\ref{poison3}(a) the parameters corresponding to the Green 
function (see inset) are $\delta =2$, $\lambda =1,$ while for the Poisson 
noise we have taken $\rho =0.3$ (dotted line), $\rho =0.4$ (dashed line) and  
$\rho =0.8$ (continuous line); in all cases taking $A=\sqrt{0.4/\rho }$. In 
consequence, the effective amplitude of the noise $\rho A^{2}$ is constant 
in the three graphs. Due to this selection the tails of the distributions 
coincide, indicating that all the distributions are asymptotically, for 
large $\mid u\mid ,$ Gaussian. This property will be analyzed analytically 
in the next-subsection. 
 
As in the case of the non-oscillating Green function, Figs.~\ref{poison1} 
and \ref{poison2}, here we expect a Gaussian stationary distribution when 
the average waiting time between two arriving pulses is shorter than the 
characteristic relaxation time of the Green function. In this case, the 
oscillatory decay of the Green function is characterized by the rate $%
\lambda /2$. Thus, we estimate that for $\rho ^{^{\prime }}>1/2$ a Gaussian 
statistics arises. This prediction agree very well with the plots shown in 
Fig.~\ref{poison3}(a). Nevertheless, note that this inequality does not 
depend on the frequency of the Green function oscillations. In the next 
figure, we will check the validity of this prediction. 
 
 
In Fig.~\ref{poison3}(b) we show the stationary distribution $P(u)$ by 
maintaining the noise parameters and changing the values of the parameters 
of the Green function. The noise parameters were chosen as $\rho =0.25$, $%
A=1 $ and for the Green function $\lambda =1$, $\delta =0.15$ (full line), $%
\delta =0.6$ (dashed line), and $\delta =50$ (dotted line). In the inset we 
show the corresponding Green function for each values. Consistently with our 
previous analysis, we confirm that independently of the value of the 
characteristic frequency of the Green function, the stationary distribution $%
P(u)$ is non Gaussian for $\rho ^{^{\prime }}<1/2$. Furthermore, we note 
that by increasing the frequency of the Green function, the fast 
oscillations of $\Lambda (t)$ completely drop-out the non-Gaussian peaks 
located at $\pm A$, remaining only one peak centered at $u=0$. On the other 
hand, with the chosen values of the noise parameters, a Gaussian 
distribution only arise for a non-oscillating Green function. 
 
Finally, we want to remark that the criteria for obtaining a stationary 
Gaussian distribution $\rho ^{^{\prime }}>\delta ^{^{\prime }}$, valid when 
the decay of the Green function is monotonous, and $\rho ^{^{\prime }}>1/2$ 
valid for an oscillatory decay, change their validity in a smooth way around 
the value $\delta ^{^{\prime }}=1/4$, which corresponds to the point in 
which the Green function modify its characteristic decay.

\subsection{Long-tail stationary distributions} 
 
Here we are interested in characterizing the stationary distribution of the 
stochastic process $u(t)$; of great importance is to know whether there will 
be or not a long-tail in the distribution. In order to make a general 
analysis, here we assume that the structure of the noise is such that we can 
write the $1-$time characteristic function in the form  
\begin{equation} 
G_{u}^{(1)}(k_{1},t_{1})=\exp \left\{ \int_{0}^{t_{1}}dt\,\,f\left( 
k_{1}\,\Lambda (t)\right) \right\} ,  \label{ST0} 
\end{equation} 
where $f\left( z\right) $ is an arbitrary function. Note that this structure 
is compatible only with a white noise. In fact, Eq.~(\ref{ST0}) follows 
using the test function $z(t)=\Theta (t_{1}-t)\,\Lambda (t_{1}-t)\,k_{1}$ in 
the general expression for the functional of $u(t)$, Eq.~(\ref{g9}), and 
assuming $G_{\xi }[k(t)]=\exp \int_{0}^{\infty }dtf(k(t))$. Taking the 
logarithm in Eq.~(\ref{ST0}), and introducing the Laplace transform (here 
denoted as ${\cal L}_{s}$) we arrive to the general expression  
\begin{equation} 
\ln G_{u}^{(1)}(k,t=\infty )=\lim_{s\rightarrow 0}s\,{\cal L}_{s}\left[ \ln 
G_{u}^{(1)}(k,t)\right] =\lim_{s\rightarrow 0}{\cal L}_{s}\left[ f\left( k%
{\bf \,}\Lambda (t)\right) \right] .  \label{ST1} 
\end{equation} 
If a Taylor expansion of $f\left( z\right) $ exists, and we can commute the 
Taylor expansion and the Laplace operator, we arrive to  
\begin{equation} 
\ln G_{u}^{st}(k)=\sum_{n=1}^{\infty }C_{n}\,k^{n}\,{\cal L}_{s}\left[ 
\Lambda (t)^{n}\right] _{s=0}.  \label{ST2} 
\end{equation} 
This formulae characterizes the stationary distribution of the stochastic 
process $u(t)$, if it exists, as a series expansion in the Fourier component  
$k$. 
 
From Eq.~(\ref{ST2}) it is simple to see that even when the stochastic 
transient can be non-Gaussian (depending on the driving noise) in the large 
asymptotic scaling $\mid u\mid \rightarrow \infty $, the Fourier transform of 
the stationary distribution goes like  
\begin{equation} 
G_{u}^{st}(k)\sim \exp \left( i\,kA_{0}-k^{2}B_{0}+\cdots \right) ,\ \text{%
for\ }k\sim 0  \label{ST3} 
\end{equation} 
where $A_{0},B_{0}$ are constants given in terms of the dissipative memory 
and the structure of the noises; then at large scale the behavior is not 
anomalous. The analysis with a non-white noise is similar and the general 
conclusion is not changed. So in order to understand the occurrence of 
anomalies or long-tails in the stationary distribution of the process $u(t)$ 
we have to consider driving noises that break the hypothesis Eq.~(\ref{ST2}%
). This case is achieved by structures like the one from the L\'{e}vy noise, 
see Eq.~(\ref{levy1}), or its associated one-side power-law noise shown in 
appendix B. 
 
In the previous sub-section we have shown that the Poisson noise, Eq.~(\ref 
{shot2}), can give rise, during the transient and in the stationary state, 
to strongly non-Gaussian distributions for the process $u(t)$. Nevertheless, 
this noise can not give rise to long-tail stationary distributions. As a 
matter of fact, the characterization of the stationary state induced by this 
shot noise follows straightforward using Eq.~(\ref{ST1}). From Eq.~(\ref 
{shot2}) its follows  
\begin{eqnarray*} 
\ln G_{u}^{st}(k) &=&\lim_{t\rightarrow \infty }\rho \int_{0}^{t}d\tau 
\,\left( \exp [iAk\Lambda (\tau )]-1\right), \\ 
&=&\frac{\rho }{\gamma }\left\{  
\mathop{\rm Ei}%
\left( iAk\right) -\ln \left( A\mid k\mid \right) -{\cal E}\right\} , 
\end{eqnarray*} 
where $\mathop{\rm Ei}(x)$ is the exponential integral function \cite{Olaf}, 
and ${\cal E}$ is the Euler constant. For simplifying the analysis we have 
assumed the local Green function Eq.~(\ref{local3}). Now it is interesting 
to remark about the non-analytic structure of $G_{u}^{st}(k)$, perhaps 
pre-announcing an anomalous behavior in the stationary distribution of the 
process $u(t)$. Nevertheless, it should be noted that in the large scale 
limit $k\rightarrow 0$ this non-analytic structure cancel out, leading 
therefore to an expansion like Eq.~(\ref{ST3}), which imply the absence of 
any long-tail distributions. This is not what happens if the driving noise 
has a long-tail structure as the one we are going to show below. 
 
\subsubsection*{Abel noise} 
 
In close connection with a Poisson noise with a random density of arriving 
pulses, in appendix B we have defined the characteristic functional of Abel 
noise ${\bf \xi }(t)$. The functional of this noise reads [$t\in (0,\infty )$%
]  
\begin{equation} 
G_{\xi }([k(t)])=\frac{2}{\Gamma (\mu )}\left( \sqrt{a\int_{0}^{\infty 
}(1-e^{ik(t)})\,dt}\right) ^{\mu }\,\,K_{\mu }\left( 2\sqrt{%
a\int_{0}^{\infty }(1-e^{ik(t)})\,dt}\right),  \nonumber 
\end{equation} 
where $\,K_{\mu }\left( z\right) $ is the Basset function and $\Gamma (\mu )$ 
the Gamma function \cite{Olaf}. Note that in contrast with the L\'{e}vy 
noise, this functional has well defined integers moments $\left\langle \xi 
(t)^{q}\right\rangle $ if $\mu >q$. Using this noise, the $1-$time 
characteristic function of the process $u(t)$ reads  
\begin{equation} 
G_{u}^{(1)}(k,t)=\frac{2\,e^{ik\left\langle u(t)\right\rangle _{0}}}{\Gamma 
(\mu )}\,\left( \sqrt{a\int_{0}^{t}(1-e^{ik\Lambda (\tau )})\,d\tau }\right) 
^{\mu }\,K_{\mu }\left( 2\sqrt{a\int_{0}^{t}(1-e^{ik\Lambda (\tau )})\,d\tau  
}\right) .  \nonumber 
\end{equation} 
From this expression (if $\Lambda (t)>0,\forall t$) it is possible to see 
that the asymptotic behavior of the probability distribution is 
characterized by a one-side power-law distribution of the Abel form  
\begin{equation} 
P(u,t)\sim \frac{{\cal A}(t)^{\mu }}{\Gamma (\mu )}\frac{\exp (-{\cal A}%
(t)/u)}{u^{1+\mu }},\ \text{for large positive }u, 
\end{equation} 
where  
\begin{equation} 
{\cal A}(t)=a\int_{0}^{t}\Lambda (\tau )\,d\tau >0.  \label{nose} 
\end{equation} 
 
In the case of local dissipation, using the Green function Eq.~(\ref{local3}%
), we get  
\begin{equation} 
{\cal A}(t)=\frac{a}{\gamma }\left( 1-e^{-\gamma t}\right) . 
\end{equation} 
From this result it is simple to see that the (one-side stable) directed 
random walk behaves like $P(u,t)\sim \left. \left( at\right) ^{\mu }\right/ 
u^{1+\mu },$ as was expected; see appendix B. 
 
In presence of a non-local dissipation and in the case $\delta >\left( \frac{%
\lambda }{2}\right) ^{2}$, due to the fact that the Green function Eq.(\ref 
{exp2}) changes its sign oscillating in time, a power-law behavior is 
obtained in a quite unusual way during the transient, i.e., the long-tail 
changes its support from a positive to a negative domain and so on as the 
time goes on. On the other hand, in the stationary regime we get  
\begin{equation} 
{\cal A}(t=\infty )=\frac{a\lambda }{\delta }, 
\end{equation} 
then also a power-law is reached at long times, with its long-tail in the 
positive domain, i.e., for $u>0,$\thinspace $P(u,t\rightarrow \infty )\sim 
\left( \frac{a\lambda }{\delta }\right) ^{\mu }\,u^{-1-\mu }.$ 
 
In Fig.~\ref{colasabelianas} we show this amazing behavior for the noise 
parameters $a=3$, $\mu =\frac{1}{2}$, $K_{1/2}(x)=~\sqrt{\pi/ 2x}\exp(-x)$.  
In this case we have used for the Green function (see inset) the parameters 
values $\delta =0.85$ and $\lambda =0.2$. At time $t=3.5$ [Fig.~\ref 
{colasabelianas}(a)] we can see that there is a long-tail in the positive 
domain, but also a sort of exponential decay appears in the negative domain 
as a result of the interplay between the oscillating Green function and the 
Abel noise. At later time $t=5$ [Fig.~\ref{colasabelianas}(b)] it is 
possible to see that the long-tail has changed its support, showing also an 
exponential decay but now in the positive domain. At time $t=50$ [Fig.~\ref 
{colasabelianas}(c)] the stationary distribution is almost reached with the 
expected long-tail in the positive domain. In these figures we have also 
fitted (with dashed lines) the corresponding long-tails with the asymptotic 
behavior  
\begin{equation} 
P(u,t)\sim \left| \frac{\Gamma (-\mu )}{\Gamma (\mu )}\right| \frac{\left|  
{\cal A}(t)\right| ^{\mu }}{\left[ sig(u)\cdot u\right] ^{1+\mu }}, 
\end{equation} 
where $sig(u)$ represents the sign of $u.$ 
 
To end this sub-section let us call the attention that by subtracting two 
statistically independent Abel noises, it is also possible to obtain the $n-$%
time characteristic function of a symmetric process $u(t)$, see appendix B. 
 
\subsection{Fractional derivative evolutions} 
 
Another example that can be characterized with the present functional 
technique is the case of fractional Langevin equations with dissipation. 
There exist many different way of introducing this kind of equations. Here 
we will consider the evolution  
\begin{equation} 
\,_{0}D_{t}^{\alpha }\left[ u\left( t\right) \right] -u_{0}\frac{t^{-\alpha }%
}{\Gamma \left( 1-\alpha \right) }=-\eta ^{\alpha }u\left( t\right) +\xi 
\left( t\right) ,  \label{fr1} 
\end{equation} 
with $1\geq \alpha >0$. This evolution was proposed to simulate the dynamics 
of financial markets where it was found that non-Gaussian driving noises 
should be the suitable ones \cite{Picozzi,Stanley}. The proper 
interpretation of this fractional stochastic equation is actually an 
integral equation  
\begin{equation} 
u\left( t\right) -u_{0}=-\eta ^{\alpha }\,_{0}D_{t}^{-\alpha 
}[u(t)]+\,_{0}D_{t}^{-\alpha }\left[ \xi \left( t\right) \right] . 
\label{fr2} 
\end{equation} 
Here, $u_{0}$ is the initial condition, and $\,_{0}D_{t}^{-\alpha }$ is the 
Riemann-Liouville integral operator \cite{Metzler}  
\begin{equation} 
\,_{0}D_{t}^{-\alpha }\left[ f\left( t\right) \right] =\frac{1}{\Gamma 
(\alpha )}\int_{0}^{t}d\tau \frac{f\left( \tau \right) }{(t-\tau )^{1-\alpha 
}}.  \label{fr3} 
\end{equation} 
It has been proved \cite{Picozzi} that the solution of this equation for 
each realization of the noises can be written as  
\begin{equation} 
u(t)=\left\langle u(t)\right\rangle _{0}+\int_{0}^{t}dt^{\prime }\,\Lambda 
(t-t^{\prime })\xi (t^{\prime }),  \label{fr4} 
\end{equation} 
where  
\begin{equation} 
\left\langle u(t)\right\rangle _{0}=u_{0}E_{\alpha ,1}[-(\eta t)^{\alpha }]. 
\label{fr5} 
\end{equation} 
On the other hand, the Green function $\Lambda \left( t\right) $\ is given 
by  
\begin{equation} 
\Lambda \left( t\right) =\Theta (t)\;t^{\alpha -1}E_{\alpha ,\alpha }[-(\eta 
t)^{\alpha }].  \label{fr6} 
\end{equation} 
Note that for this model of dissipative fractional equations the average $%
\left\langle u(t)\right\rangle _{0}$ and the noise propagate with different 
Green functions. The generalized Mittag-Leffler function $E_{\alpha ,\beta }$ 
is defined by  
\begin{equation} 
E_{\alpha ,\beta }\left( z\right) =\sum_{k=0}^{\infty }\frac{z^{k}}{\Gamma 
\left( \alpha k+\beta \right) }\;\;\;\alpha >0,\beta >0.  \label{fr7} 
\end{equation} 
From Eq.~(\ref{fr4}) it is evident that our functional approach allows us to 
introduce any kind of statistics to drive the fractional stochastic 
evolution. 
 
\subsubsection*{Competition between fractional derivative and L\'{e}vy noise} 
 
Here we are interested in analyzing the competence between the statistic of 
L\'{e}vy noise and the dynamical effects introduced by a fractional 
derivative structure. In order to get a simpler analysis, here we will 
assume zero dissipation, i.e., $\eta =0$. In this case, the previous 
evolution reduce to  
\begin{equation} 
\left\langle u(t)\right\rangle _{0}=u_{0},\;\;\;\;\;\;\;\;\;\;\;\;\Lambda 
\left( t\right) =\frac{\Theta (t)}{\Gamma (\alpha )}\frac{1}{t^{1-\alpha }}. 
\label{marinera} 
\end{equation} 
Note that from differentiation of Eq.~(\ref{fr2}), these two previous 
expressions characterizes the evolution  
\begin{equation} 
\frac{du(t)}{dt}=\,_{0}D_{t}^{1-\alpha }\xi (t).  \label{tetera} 
\end{equation} 
We would like to stress that this problem in the frame of our functional 
approach can be done in a simple way. Using Eq.~(\ref{marinera}) and the 
functional of a L\'{e}vy white noise, Eq.~(\ref{levy1}), the process $u(t)$ 
is fully characterized by the function [see Eq.~(\ref{Levymemo})]  
\begin{equation} 
\Xi _{u}(t)=\sigma _{\nu }\left( \frac{1}{\Gamma (\alpha )}\right) ^{\nu 
}\,\left( \frac{1}{1-\nu (1-\alpha )}\right) t^{1-\nu (1-\alpha )}. 
\end{equation} 
This expression is only well defined if $\nu (1-\alpha )<1$. Note the 
restriction on the parameters $\{\nu ,\alpha \}$. This result contrasts with 
that of a driving Gaussian white noise, where the restriction is $\alpha \in 
(\frac{1}{2},1]$. Thus, in the case of L\'{e}vy noise, smaller values of the 
parameter $\alpha \in (0,1]$ are allowed at the expenses of diminishing the 
value of $\nu \in (0,2]$ (i.e., large L\'{e}vy-step excursions). 
 
Finally, we want to remark that Eq.~(\ref{tetera}) driven by Gaussian 
fluctuations, can be mapped with an effective medium approximation in the 
context of disordered systems \cite{libro,desorden}. Note that from our 
functional approach, it is easy to calculate any correlation functions and 
more complicated objects than the propagator of the system; this is 
something that is hard to obtain in the context of self-consistent 
approximations \cite{ABOS73}. 
 
\section{Summary and Conclusions} 
 
We have completely characterized generalized linear Langevin equations when 
the usual fluctuation-dissipation theorem does not apply. Our central result 
is an exact expression for the characteristic functional of the process $%
u(t) $, for a general $d-$dimensional correlated process (in appendix A we 
present a $2-$dimensional example) where the memory and the noise are 
arbitrary. In this way we have been able to give a closed expression to get 
the whole Kolmogorov hierarchy, i.e., to calculate any $n-$joint probability 
and any $n-$time correlation function or cumulant. 
 
We have applied our formalism to many different noise structures, and in 
this way we have shown the noise-induced interplay between the effects of 
the dissipative memory and the structure of the driving noises. Particular 
emphasis have been put in analyzing the $1-$time probability distribution $%
P(u,t)$ and its stationary state; we have shown that the transient toward 
the stationary state strongly depends on the interplay between the 
dissipation and the noise structure. As example we have analyzed the case of 
an exponential memory function in presence of different stochastic driving 
forces; for example, radioactive noise and Poisson noise. Also we have 
discussed whether or not, at large scale, a non-Gaussian distribution 
appears and when these distributions have long-tails. In order to get this 
class of stationary distributions we have introduced, for the first time, 
the Abel noise in connection with the occurrence of one-side power-law 
distributions (see appendix B). We remark that this noise structure can be 
useful when studying models where molecular diffusion is an important 
ingredient to be considered. In this context, the relaxation analysis of a 
generalized Langevin particle considering memory dissipation and injection 
of energy by microscopic random contributions characterized by a symmetric 
power-law, can be also carried out by using the present functional 
formalism. As a matter of fact, asymptotically when the Abel noise is 
symmetric the conclusions, for $\mu <2$, are in agreement with the structure 
of the L\'{e}vy noise. 
 
Of particular relevance is the use of our formalism to study fractional 
Langevin equations driven by arbitrary noise structures. We have analyzed 
the competence between the noise statistic and the fractional derivative 
operator. As examples we have compared L\'{e}vy and Gaussian noises and 
their interplay with the fractional calculus. We remark that the possibility 
of introducing arbitrary noise statistics in fractional Langevin equations 
is an interesting step forward to broad the possible applications of these 
equations \cite{Picozzi}. 
 
To end this work let us comment that the present functional approach can 
also be applied to the so called delayed Langevin equations. In that case 
the associated Green function $\Lambda (t)$ turns to be the crucial 
ingredient to study different models of delayed Langevin equations driven by 
arbitrary noises. Results along this line will be presented elsewhere. 
 
\appendix  
 
\section{A Two Dimensional Example} 
 
In what follows we will show a multivariable example that can be fully 
worked out by using our formalism. It corresponds to a particle confined in 
an harmonic potential and driven by an arbitrary noise $\xi (t)$ 
(fluctuating force). The evolution for the position $x(t)$\ and the velocity  
$v\left( t\right) $ reads  
\begin{eqnarray} 
\frac{d}{dt}x(t) &=&v\left( t\right),  \label{ho1} \\ 
\frac{d}{dt}v(t) &=&-\Omega ^{2}x(t)-\int_{0}^{t}dt^{\prime }\,\Phi \left( 
t-t^{\prime }\right) v(t^{\prime })+\xi (t),  \label{ho2} 
\end{eqnarray} 
where $\Omega $ is the characteristic frequency of the harmonic potential. 
All statistical information of the processes $[x(t),v\left( t\right) ]$ is 
encoded in the characteristic functional  
\begin{equation} 
G_{xv}([k_{x}(t),k_{v}(t)])=\left\langle \exp i%
\textstyle \int%
\limits_{0}^{\infty }dt\,\left( k_{x}(t)x(t)+k_{v}(t)v(t)\right) 
\right\rangle .  \label{ho3} 
\end{equation} 
For each realization of the noise, the solution of Eqs.~(\ref{ho1}) and (\ref 
{ho2}) reads  
\begin{eqnarray} 
x(t) &=&\left\langle x(t)\right\rangle _{0}+\int_{0}^{t}dt^{\prime }\,\,%
{\frak X}(t-t^{\prime })\xi (t^{\prime }),  \label{ho4} \\ 
v(t) &=&\left\langle v(t)\right\rangle _{0}+\int_{0}^{t}dt^{\prime }\,{\frak %
F}\left( t-t^{\prime }\right) \xi (t^{\prime }).  \label{ho5} 
\end{eqnarray} 
Here, we have defined the averages  
\begin{eqnarray} 
\left\langle x(t)\right\rangle _{0} &=&x(0)+\int_{0}^{t}dt^{\prime 
}\,\left\langle v(t^{\prime })\right\rangle _{0}  \label{ho6} \\ 
\left\langle v(t)\right\rangle _{0} &=&v\left( 0\right) {\frak F}\left( 
t\right) -\Omega ^{2}x(0)\,{\frak X}(t).  \label{ho7} 
\end{eqnarray} 
The functions ${\frak F}\left( t\right) $ and ${\frak X}(t)$\ are defined 
from their Laplace transforms  
\begin{equation} 
{\frak X}(s)=\frac{{\frak F}\left( s\right) }{s},\ {\frak F}\left( s\right) =%
\frac{s}{s\left[ s+\Phi \left( s\right) \right] +\Omega ^{2}},  \label{ho8} 
\end{equation} 
which is equivalent to  
\begin{eqnarray} 
\frac{d}{dt}{\frak X}(t) &=&{\frak F}(t),  \label{ho9} \\ 
\frac{d}{dt}{\frak F}(t) &=&-\Omega ^{2}\,{\frak X}(t)-\int_{0}^{t}dt^{%
\prime }\,\Phi \left( t-t^{\prime }\right) {\frak F}(t^{\prime }), 
\label{ho10} 
\end{eqnarray} 
with the initial conditions ${\frak X}(0)=0$ and ${\frak F}(0)=1$. After 
inserting Eqs.~(\ref{ho4})-(\ref{ho5}) in the definition Eq.~(\ref{ho3}) we 
arrive to the exact expression  
\begin{equation} 
G_{xv}([k_{x}(t),k_{v}(t)])=G_{\left\langle x\right\rangle 
_{0}}([k_{x}(t)])\,G_{\left\langle v\right\rangle _{0}}([k_{v}(t)])\,G_{\xi 
}([z(t)]),  \label{ho11} 
\end{equation} 
where we have defined  
\begin{eqnarray} 
G_{\left\langle x\right\rangle _{0}}([k_{x}(t)]) &=&\exp \left\{ i%
\textstyle \int%
_{0}^{\infty }dt\,k_{x}(t)\left\langle x(t)\right\rangle _{0}\right\} , 
\label{ho12} \\ 
G_{\left\langle v\right\rangle _{0}}([k_{v}(t)]) &=&\exp \left\{ i%
\textstyle \int%
_{0}^{\infty }dt\,k_{v}(t)\left\langle v(t)\right\rangle _{0}\right\} . 
\label{ho13} 
\end{eqnarray} 
Furthermore, the scalar function $z(t)$\ is defined by  
\begin{equation} 
z(t)=\int_{t}^{\infty }dt^{\prime }\,{\frak X}(t^{\prime }-t)k_{x}(t^{\prime 
})+\int_{t}^{\infty }dt^{\prime }\,{\frak F}(t^{\prime }-t)k_{v}(t^{\prime 
}).  \label{ho14} 
\end{equation} 
Thus, knowing the characteristic functional of the noise allows to obtain 
the full characteristic functional of the bidimensional process $\left[ 
x\left( t\right) ,v\left( t\right) \right] $. 
 
{\it Free Brownian motion}: By taking $\Omega =0$\ , the previous case 
reduces in a trivial way to the case of a free Brownian particle. In this 
case the evolution of the particle position is given by  
\begin{equation} 
\frac{d^{2}x(t)}{dt^{2}}=-\int_{0}^{t}dt^{\prime }\,\Phi \left( t-t^{\prime 
}\right) \,\frac{dx(t^{\prime })}{dt^{\prime }}+\xi (t). 
\end{equation} 
This second order differential equation also arises when modeling a rigid 
rotator \cite{Caceres,Manuel}. The marginal statistic description of this 
equation can be obtained from Eq.~(\ref{ho11}) after taking $k_{v}(t)=0$. 
Quantities like $\left\langle \cos [x(t)]\right\rangle $ or $\left\langle 
\sin [x(t)]\right\rangle $ follow immediately from the real and imaginary 
part of the characteristic functional, and more complex stochastic objects 
can also be calculated analytically. In general, any order differential 
stochastic equation can be analyzed with our functional formalism after 
introducing additional stochastic process corresponding to the different 
derivatives of the original process. 
 
\section{Abel noise} 
 
Abel was probably the first to give an application of fractional calculus  
\cite{Abel}. He used derivatives of arbitrary order to solve the isochrone 
problem in classical mechanics, and the integral equation he worked out was 
precisely the one Riemann used to define fractional derivatives. In a modern 
context, the Abel-type of integral equation can be written in the form  
\begin{eqnarray} 
\xi ^{2\mu }\,P(\xi ) &=&a^{\mu }\,\frac{1}{\Gamma (\mu )}\int_{0}^{\xi 
}\left( \xi -y\right) ^{\mu -1}P(y)\,dy,\;\;\;\;\; \mu >0,  \label{A1} \\ 
&=&a^{\mu }\,_{0}D_{\xi }^{-\mu }\,P(\xi ),  \nonumber 
\end{eqnarray} 
where $a$ is an arbitrary constant. It is interesting to note that the 
particular class of normalized one-side L\'{e}vy-type of probabilities  
\begin{equation} 
P(\xi )=\frac{a^{\mu }}{\Gamma (\mu )}\xi ^{-1-\mu }\exp \left( -a/\xi 
\right) ,\ \text{with}\;\; a>0, \; \xi >0,  \label{A2} 
\end{equation} 
are solution of the fractional differential equation Eq.~(\ref{A1}). The 
characteristic function of the probability density Eq.~(\ref{A2}) can be 
calculated taking the Laplace transform, then  
\begin{eqnarray} 
G_{\xi }(k) &=&\left\langle \exp \left( ik\xi \right) \right\rangle ={\cal L}%
_{s}\left[ P(\xi )\right] _{s=-ik}  \nonumber \\ 
&=&\frac{2}{\Gamma (\mu )}\,\left( \sqrt{-ika}\right) ^{\mu }\,K_{\mu 
}\left( 2\sqrt{-ika}\right) , 
\end{eqnarray} 
where $K_{\mu }(z)$ is the Basset function and $\Gamma (\mu )$ is the Gamma 
function. Note that if $\mu \in (0,1)$ the asymptotic behavior of the Basset 
function shows the expected divergencies for integers moments $\left\langle 
\xi ^{q}\right\rangle $, but in general depending on the value of $\mu $ we 
can get finite moments  
\begin{equation} 
\left\langle \xi ^{q}\right\rangle =a^{q}\,\frac{\Gamma (\mu -q)}{\Gamma 
(\mu )},\quad \text{for}\quad \mu >q. 
\end{equation} 
We define the characteristic functional of a stochastic process $\xi (t)$, 
closely related to the power-law distribution Eq.~(\ref{A2}), in the form  
\begin{equation} 
G_{\xi }([k(t)])=\frac{2}{\Gamma (\mu )}\left( \sqrt{a\int_{0}^{\infty 
}(1-e^{ik(t)})\,dt}\right) ^{\mu }\,\,K_{\mu }\left( 2\sqrt{%
a\int_{0}^{\infty }(1-e^{ik(t)})\,dt}\right) .  \label{A3} 
\end{equation} 
From this expression all the moments of the noise can be calculated. For 
example the first moments read  
\begin{eqnarray} 
\left\langle \xi (t)\right\rangle &=&a\frac{\Gamma (\mu -1)}{\Gamma (\mu )}%
,\;\; \text{if}\;\; \mu >1,  \label{A30} \\ 
\left\langle \xi (t_{1})\xi (t_{2})\right\rangle &=&a\frac{\Gamma (\mu -1)}{%
\Gamma (\mu )}\,\delta (t_{1}-t_{2})+a^{2}\frac{\Gamma (\mu -2)}{\Gamma (\mu 
)},\; \text{if}\;\; \mu >2.  \nonumber 
\end{eqnarray} 
Note that there is a constant term in the correlation function $\left\langle 
\left\langle \xi (t_{1})\xi (t_{2})\right\rangle \right\rangle \equiv 
\left\langle \xi (t_{1})\xi (t_{2})\right\rangle -\left\langle \xi 
(t_{1})\right\rangle \left\langle \xi (t_{2})\right\rangle $,  
\begin{equation} 
\left\langle \left\langle \xi (t_{1})\xi (t_{2})\right\rangle \right\rangle =%
\frac{a}{(\mu -1)}\delta (t_{1}-t_{2})+\left( \frac{a}{(\mu -1)}\right) ^{2}%
\frac{1}{(\mu -2)},\ \text{if}\ \mu >2.  \label{A31} 
\end{equation} 
As expected this constant term decreases for large values of $\mu $. From 
now on we will call $\xi (t)$ the Abel noise in honor to that brilliant 
mathematician. 
 
Now in order to have a clear meaning of this noise we introduce an 
alternative interpretation. Let a stochastic process $n(t)$ be defined by a 
directed random walk, then for a given value of transition rate $\rho $ we 
write the master equation  
\begin{equation} 
\frac{dP(n,t)}{dt}=\rho \,[P(n-1,t)-P(n,t)],\quad \rho >0.  \label{A33} 
\end{equation} 
Using the Darling-Siegert theorem \cite{Van} the characteristic functional 
of the process $n(t)$ can be found by solving the masterly equation  
\begin{equation} 
\frac{d\Pi (n,t)}{dt}=ik(t)n\Pi (n,t)+\rho \left[ \Pi (n-1,t)-\Pi 
(n,t)\right] ,\quad \Pi (n,0)=\delta (n-n_{0}),  \label{A4} 
\end{equation} 
where $k(t)$ is any test function. The characteristic functional of the 
process $n(t)$ follows from the limit  
\begin{equation} 
G_{n}([k(t)])=\lim_{t\rightarrow \infty }\sum_{n=0}^{\infty }\Pi (n,t). 
\label{A40} 
\end{equation} 
Using the generating function $\Pi (Z,t)=\sum_{n=0}^{\infty }Z^{n}\,\Pi 
(n,t),$ the solution of the generating function of Eq.~(\ref{A4}) is, with 
the abbreviation $\int_{0}^{t}k(t^{\prime })\,dt^{\prime }=K(t)$,  
\begin{equation} 
\Pi (Z,t)=\left( Ze^{iK(t)}\right) ^{n_{0}}\,\exp \left( \int_{0}^{t}\rho 
\left[ Ze^{iK(t)-iK(t^{\prime })}-1\right] \,dt^{\prime }\right) , 
\end{equation} 
so using Eq.~(\ref{A40}) we finally arrive to the functional  
\begin{equation} 
G_{n}([k(t)])=e^{iK(\infty )\,n_{0}}\,\exp \left( \rho \int_{0}^{\infty 
}\left[ e^{i\int_{t}^{\infty }k(t^{\prime })\,dt^{\prime }}-1\right] 
\,dt\right) .  \label{A5} 
\end{equation} 
From this expression it is simple to see that  
\begin{equation} 
G_{n}([k(t)])=e^{iK(\infty )\,n_{0}}\,G_{\xi }([z(t)]), 
\end{equation} 
where $G_{\xi }([z(t)])$ is the functional of the Poisson noise ($A=1$) 
evaluated in the test function $z(t)=\int_{t}^{\infty }k(t^{\prime 
})\,dt^{\prime }$. This means that the Markov process $n(t)$ controlled by 
Eq.~(\ref{A33}), with a probability rate $\rho $, is equivalent to solve a 
linear Langevin equation driven by a Poisson noise $\xi (t)$ with a uniform 
density of arriving pulses $\rho $. Thus we can write the associated 
stochastic differential equation  
\begin{equation} 
\frac{dn}{dt}=\xi (t).  \label{A6} 
\end{equation} 
Now, we will assume that the rate $\rho $ in Eq.~(\ref{A33}) is a random 
variable with a distribution $P(\rho )$. This assumption arises naturally in 
the context of disordered systems. Therefore, the final functional can be 
obtained from Eq.~(\ref{A5}) as  
\begin{equation} 
\left\langle G_{n}([k(t)])\right\rangle _{P(\rho )}=e^{iK(\infty 
)\,n_{0}}\,\int_{0}^{\infty }G_{\xi }([z(t)])\,P(\rho )\,d\rho . 
\end{equation} 
This integral can be done for many different distributions $P(\rho ),$ in 
particular if we use the Abel distribution Eq.~(\ref{A2}), we get  
\begin{equation} 
\int_{0}^{\infty }G_{\xi }([z(t)])\,P(\rho )\,d\rho =\frac{2}{\Gamma (\mu )}%
\left( \sqrt{a\int_{0}^{\infty }(1-e^{iz(t)})\,dt}\,\right) ^{\mu }\,K_{\mu 
}\left( 2\sqrt{a\int_{0}^{\infty }(1-e^{iz(t)})\,dt\,}\right) , 
\end{equation} 
where we have used that ${\cal R}_{e}\left[ \int_{0}^{\infty 
}(e^{iz(t)}-1)\,dt\right] \leq 0$. This result ends the interpretation of 
the Abel noise as the noisy term appearing in Eq.~(\ref{A6}), in close 
connection with the solution of a directed random walk model with a random 
(power-law distributed) transition rate $\rho $. 
 
Note that the cumulants of the Abel noise are not self-averaging with 
respect to the cumulants of the Poisson noise and the average over $P(\rho )$%
. For example using the Poisson functional, see Eq.~(\ref{shot2}) with $A=1$ 
[or Eq.~(\ref{A5})], it is simple to show that its second cumulant is $%
\left\langle \left\langle \xi (t_{1})\xi (t_{2})\right\rangle \right\rangle 
=\rho \,\delta (t_{1}-t_{2})$. Then the average of the correlation of the 
Poisson noise reads  
\begin{equation} 
\int_{0}^{\infty }d\rho \ P(\rho )\left\langle \left\langle \xi (t_{1})\xi 
(t_{2})\right\rangle \right\rangle =a\frac{\Gamma (\mu -1)}{\Gamma (\mu )}%
\,\delta (t_{1}-t_{2}), 
\end{equation} 
result that is different from the calculation of the correlation of the Abel 
noise, see Eq.~(\ref{A31}). 
 
Consider now the subtraction of two Abel noises $\xi ^{c,b}(t)$ statistical 
independent: $\xi (t)=\xi ^{c}(t)-\xi ^{b}(t)$. In this case, we can write 
the characteristic functional of the process $\xi (t)$ in the form  
\begin{eqnarray} 
G_{\xi }([k(t)]) &=&\left\langle \exp \left( i\int_{0}^{\infty }k(t)\left[ 
\xi ^{c}(t)-\xi ^{b}(t)\right] \,dt\right) \right\rangle  \nonumber \\ 
&=&G_{\xi ^{c}}([k(t)])\,G_{\xi ^{b}}([-k(t)]), 
\end{eqnarray} 
where each functional is given by Eq.~(\ref{A3}). From the properties of the 
Basset function we can write this formula in a compact form, using the 
Kelvin functions, to handle this in a more friendly way. Also from this 
expression, it is simple to see that the first moment of the symmetric Abel 
noise $\xi (t)$ is null, etc.


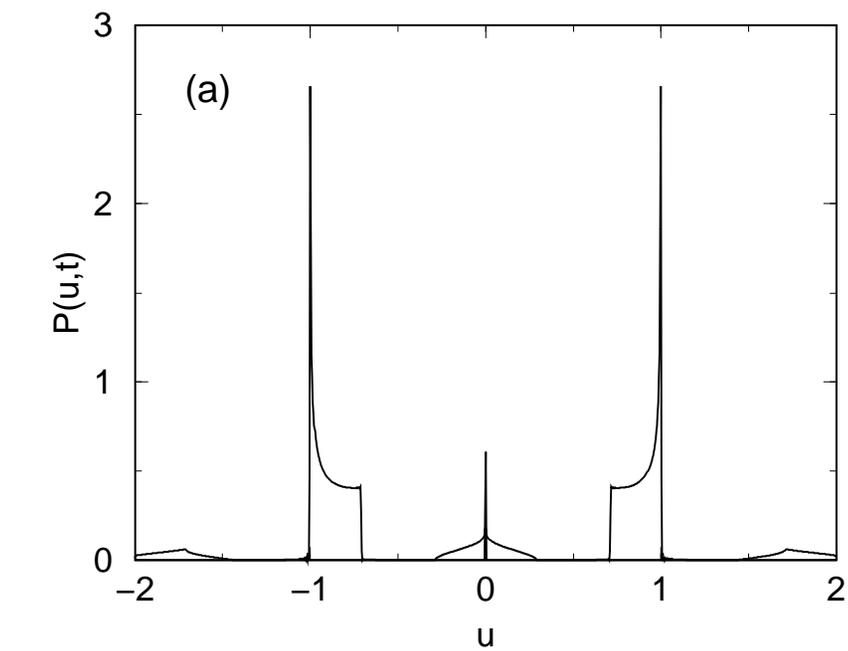
\begin{figure}[tbp]  
\caption{ 
Probability distribution $P(u,t)$ as in Fig.~\ref{poison1}, for four 
different times, Fig.~\ref{poison2}(a) $t=2.5$, Fig.~\ref{poison2}(b) $t=5$, 
Fig.~\ref{poison2}(c) $t=10$, and Fig.~\ref{poison2}(d) $t=100$ where the 
stationary state is practically reached. The noise parameters are $A=1$ and $%
\rho =0.1$. The parameters of the Green function are as in Fig.~\ref{poison1}%
; the inset in Fig.~\ref{poison2}(d) shows its monotonous decaying behavior 
as a function of time. The $\delta -$Dirac contributions (see text) are 
indicated with straight lines. 
} 
\label{poison2}  
\end{figure}


\begin{figure}[tbp]  
\caption{ 
Probability distribution, $P(u,t),$ for a memory-like driven process $u(t)$ 
with an exponential dissipative kernel and in presence of the Abel noise, $%
\xi (t),$ as a function of $u$ for three different times in arbitrary units, 
Fig.~\ref{colasabelianas}(a) $t=3.5$, Fig.~\ref{colasabelianas}(b) $t=5$ , 
Fig.~\ref{colasabelianas}(c) $t=50$. The initial condition was $u(t=0)=0$. 
The noise parameters are $a=3$ and $\mu =\frac{1}{2}$. The parameters of the 
Green function are $\delta =0.85$ and $\lambda =0.2$; the inset shows its 
oscillatory decaying behavior as a function of time. The dashed lines 
correspond to the long-tail fit (see text). In Fig.~\ref{colasabelianas}(c) 
the stationary regime of the distribution of the process $u(t)$ has been 
reached.
} 
\label{colasabelianas}  
\end{figure}

 
 
 
 
 


\begin{references}
\bibitem{desorden}  This mapping can be done through the corresponding 
generalized diffusion coefficient $D_{\alpha }(s)=(s^{2}/2){\cal L}%
_{s}\left[ \left\langle \left\langle u(t)u(t)\right\rangle \right\rangle 
\right] $, which in the present case reads  
\[ 
D_{\alpha }(s)=\left( \frac{\kappa _{\alpha }}{\Gamma (\alpha )}\right) 
^{2}\left( \frac{\Gamma (2\alpha )}{2\alpha -1}\right) \frac{\,s^{2(1-\alpha 
)}}{2},\ \alpha \in (\frac{1}{2},1]. 
\] 
On the other hand from an effective medium approximation \cite{ABOS73}, the 
self-consistent Laplace-dependent diffusion coefficient in $1D$ is given, 
asymptotically, by  
\[ 
D(s)\approx \left( \frac{\sin \pi \zeta }{2\pi (1-\zeta )}\right) 
^{2/(2-\zeta )}\ \ s^{\zeta /(2-\zeta )}. 
\] 
Thus, it is simple to see that in the long time limit ($s\sim 0$) there is 
map between the fractional index $\alpha $ and the exponent $\zeta $ that 
characterizes the strength of the disorder  
\[ 
\alpha =\frac{4-3\zeta }{4-2\zeta }\Rightarrow \,\alpha \in (\frac{1}{2},1]. 
\] 
Note that our functional technique also allows us to calculate more 
complicated objects in disordered systems; for example, we can calculate the 
non trivial two-particles density $P(u_{1},t_{1};u_{2},t_{2})$, etc.
\end{references}
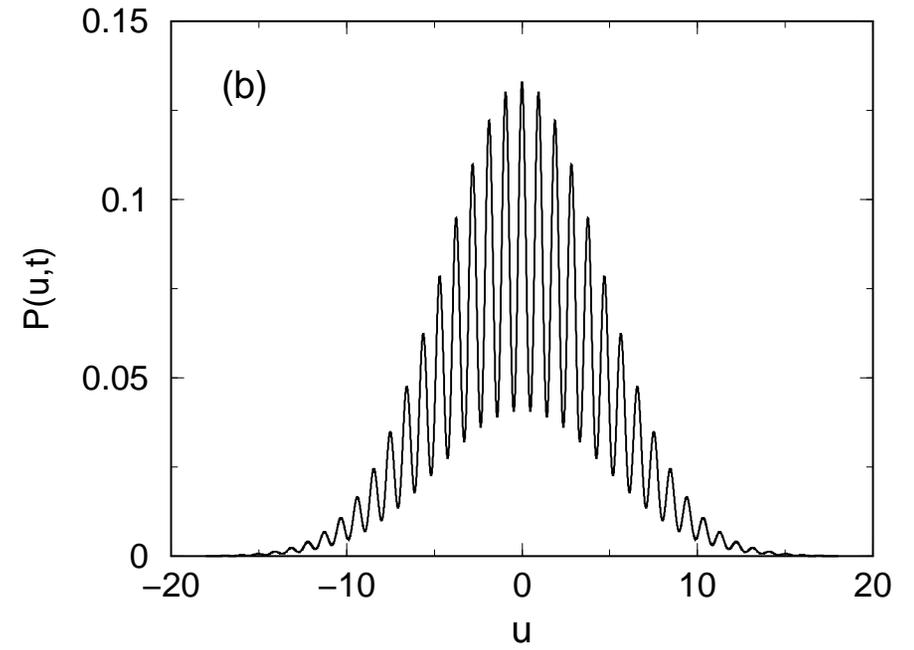
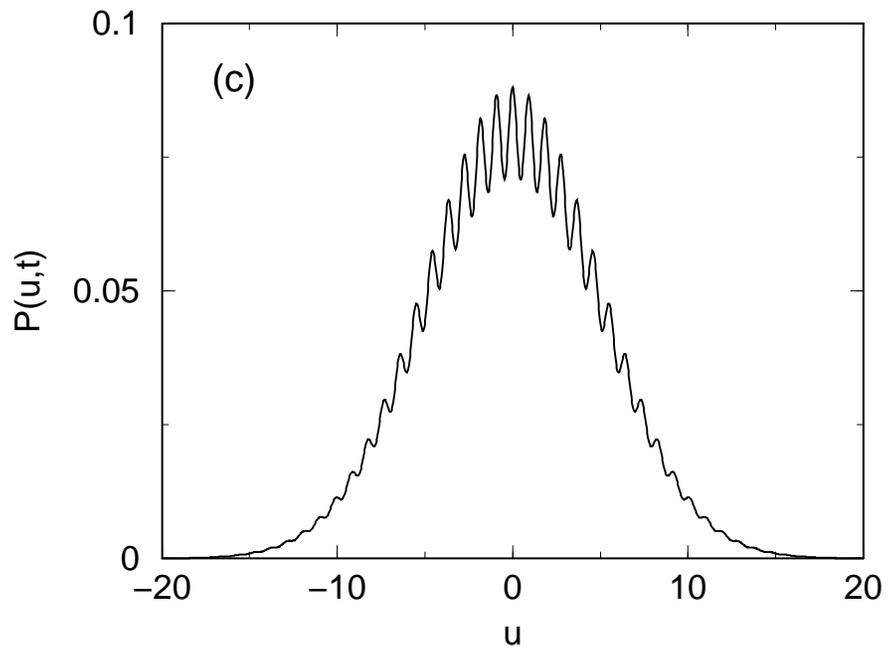
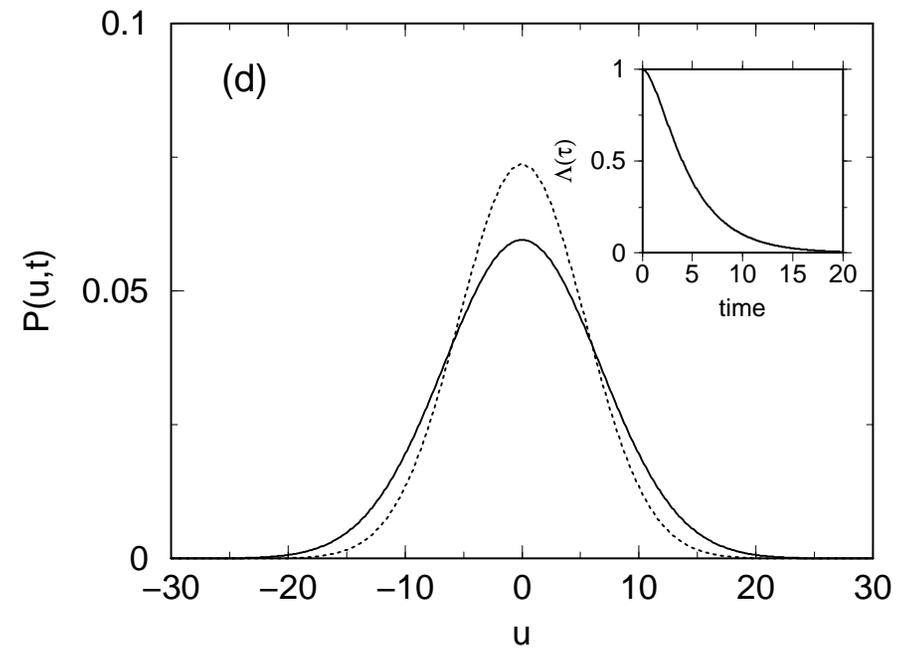
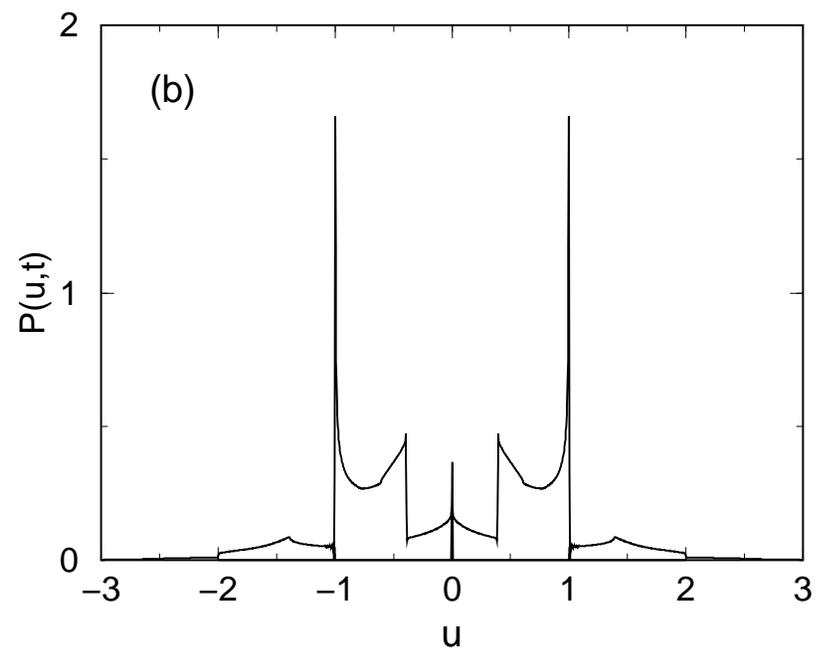
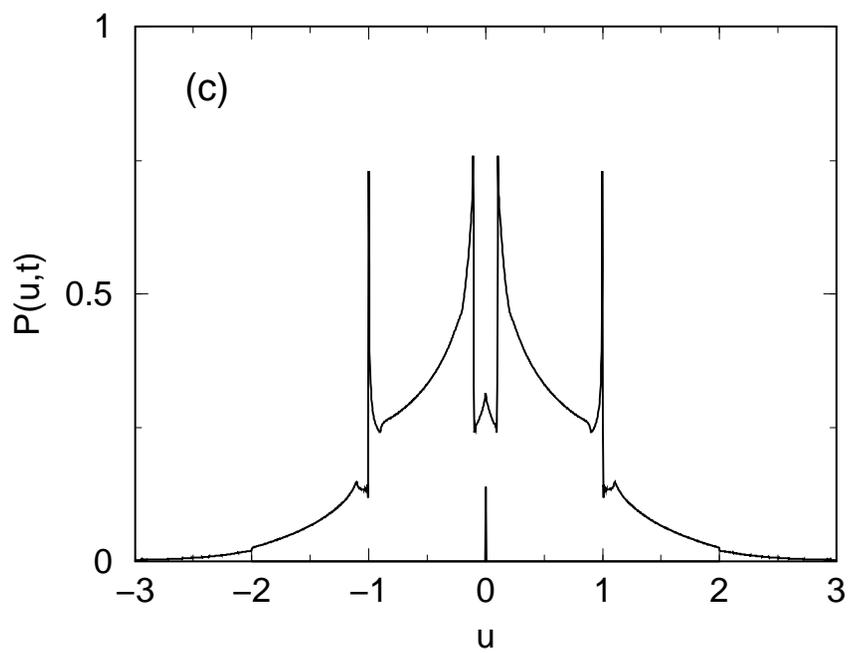
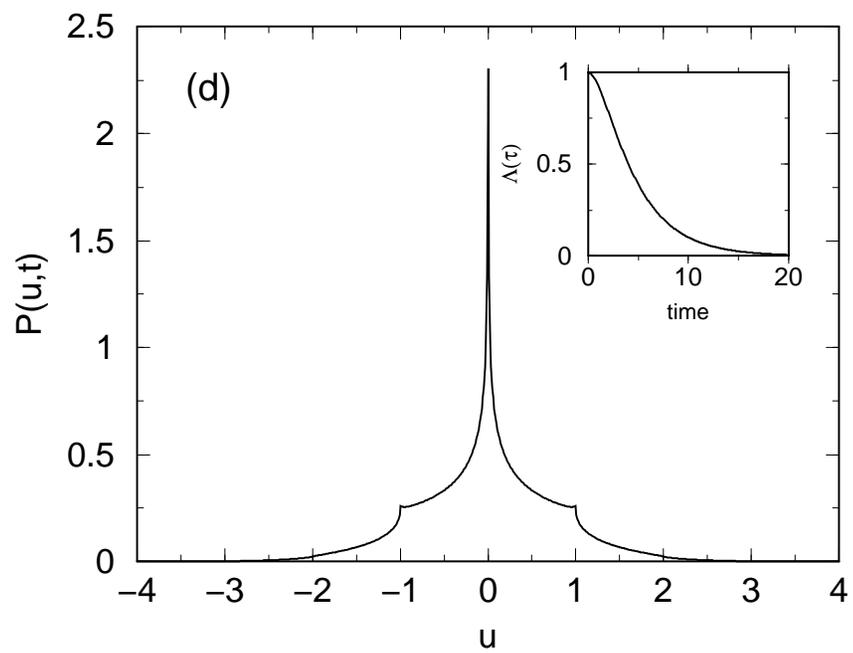
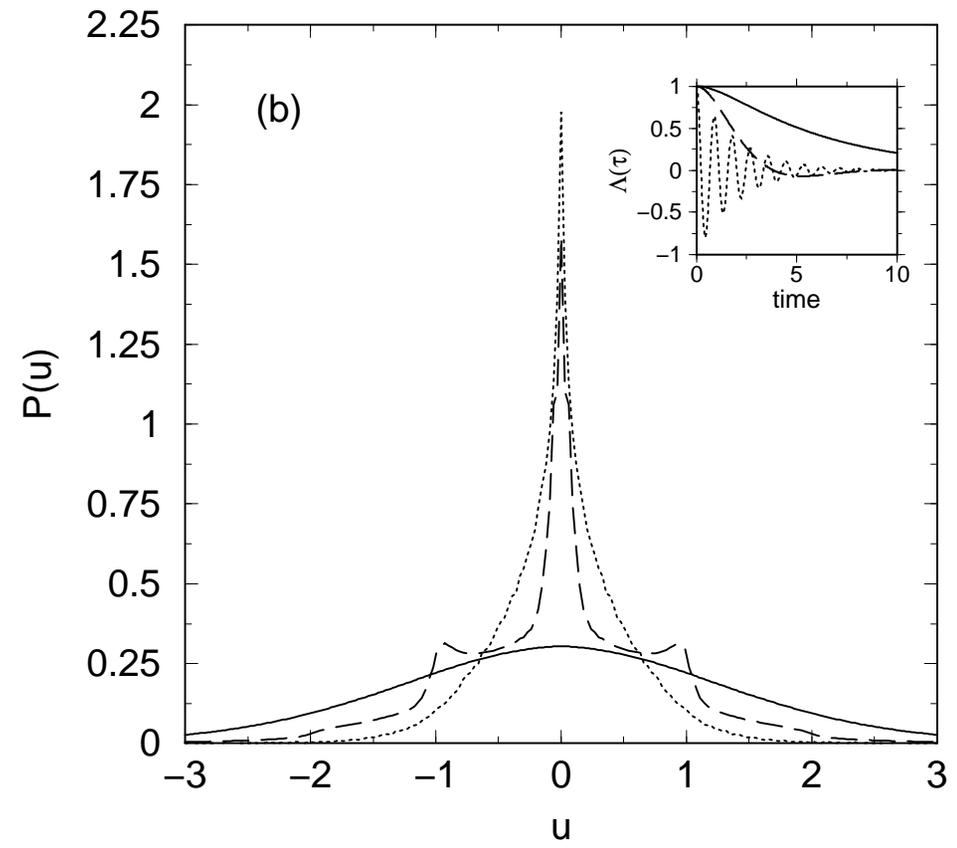
\end{document}